%% file: main.tex
\documentclass[sigconf]{acmart}
\usepackage{booktabs} 
\usepackage{multirow}
\usepackage{flushend}
\usepackage{subfig}
\usepackage{listings}
\usepackage[utf8]{inputenc}

\begin{document}
\fancyhead{}
\title{An Empirical Study of Blockchain-based Decentralized Applications}

\author{Kaidong Wu$^1$}
\affiliation{%
  \institution{$^1$Key Lab of High-Confidence Software Technology, MoE (Peking University), Beijing, China\\
  }
}
\email{wukd94@pku.edu.cn}

\renewcommand{\shortauthors}{K. Wu et al.}

\begin{abstract}
\input{secs/abstract}
\end{abstract}




\keywords{Decentralized Applications; Ethereum; Blockchain; Empirical Study}

\maketitle

\input{secs/introduction}
\input{secs/background}
\input{secs/dataset}
\input{secs/descriptive}
\input{secs/smartcontract}
\input{secs/findings}
\input{secs/implications}
\input{secs/relatedwork}
\input{secs/conclusion}
\bibliography{main}{}
\bibliographystyle{ACM-Reference-Format}

\end{document}

%% file: secs/abstract.tex
A decentralized application (dapp for short) refers to an application that is executed by multiple users over a decentralized network. In recent years, the number of dapp keeps fast growing, mainly due to the popularity of blockchain technology.  Despite the increasing importance of dapps as a typical application type that is assumed to promote the adoption of blockchain, little is known on what, how, and how well dapps are used in practice. In addition, the insightful knowledge of whether and how a traditional application can be transformed to a dapp is yet missing. To bridge the knowledge gap, this paper presents a comprehensive empirical study on an extensive dataset of 734 dapps that are collected from three popular open dapp marketplaces, i.e., ethereum, state of the dapp, and DAppRadar. We analyze the popularity of dapps, and summarize the patterns of how smart contracts are organized in a dapp. Based on the findings, we draw some implications to help dapp developers and users better understand and deploy dapps.

%% file: secs/introduction.tex
\section{Introduction}

Nowadays blockchain has become a hot topic in both industry and academia. As the first cpytocurrency and blockchain application, Bitcoin started in Jan, 2009, and its capital market reaches over 100 billion dollars now. Blockchain can be regarded as a public ledger and all committed transactions are stored in a list of blocks. The chain grows as new blocks are appended to it continuously. Bitcoin is seen as the first generation of blockchain compared to Ethereum. Ethereum started at May 30, 2015, and proposes a new mechanism called smart contract: users deploy them to blockchain and then they can be called anywhere. Results they generate are stored in blockchain. Decentralized applications (dapps) appear based on these features.

Ethereum dapps have become a market of billions of dollars and millions of users. A huge volume of data have been generated by usage of dapps. Understanding which kind of dapps are popular is meaningful to all stakeholders. For example, dapp collecting website owners can guide web traffic by related dapp list of popular dapps and provide effective dapp recommendation services; developers can understand how to use smart contracts provide services, why their dapps are popular or not, and improve the design; end-users can have a better knowledge about dapps and use them better. Due to the lack of related researches, most of these important questions don't have satisfactory answers.

Endless reports of websites have shown some features of dapp market. Most of these analyses were conducted using transaction data from some open dapp collection websites. They only focus on daily active users, weekly transactions, transaction volume of a month and so on. In academia, there are some studies about usage of Bitcoin, Ethereum and smart contracts. But few studies focus on dapps' transaction data, smart contracts, LoC and systematic descriptive analysis based on those above.

We analyze data collected from 734 dapps and 2,740 smart contracts. Out data are from two open websites collecting dapps and a Ethereum block explorer website. They provide detailed data about dapps, transactions, smart contracts and source codes.

To the best of our knowledge, our work is the first descriptive analysis of dapp. Based on such a dataset, we get some conclusions and insights, and suggest some implications. In total, our study explores the following questions:

\begin{itemize}
    \item \textbf{How can we characterize the dapp popularity?} By dapps' distribution by categories, we find that difference between 16 categories of dapps. Then we explore the dapp popularity by unique users, transactions and transaction volumes, and find that distributions of all these aspects follow the Pareto principle, namely a few dapps have substantial popularity. These findings can help dapp collecting website owners improve their advertising and recommendation services.
    \item \textbf{Does open source influence the popularity of dapps?} We judge whether a dapp is open-source by their off-chain codes and smart contract source codes. In our dataset, all open-source dapps are open-sourced by Github \cite{github} and Etherscan \cite{etherscan} (introduced in Section \ref{sec:etherscan}). We find that open source of smart contract source codes influence popularity of dapps. It will help developers open their smart contract source codes.
    \item \textbf{Is there reuse in dapps?} We explore reuse in dapps. We find code-level reuse is common in dapps, and used for different purposes. These findings can help developers rationally design their dapps, and warn users to check source codes when using dapps of category Gambling.
    \item \textbf{How do developers transfer an app to a dapp?} We select a hot dapp and analyze its source code. By case study, we find smart contracts are usually used to hold token related data. Just a few operations can be transferred to smart contracts. Then we conclude how to do the transformation.
\end{itemize}

The remaining of this paper is organized as follows. We introduce related technologies of dapps, including blockchain and smart contract in Section \ref{sec:background}. Section \ref{sec:dataset} describes our dataset. In Section \ref{sec:descriptive} we do some descriptive analysis, and then do in-depth analysis in Section \ref{sec:sc}. Section \ref{sec:findings}-\ref{sec:implications} describe our findings and implications. Section \ref{sec:related_work} summarizes the related work, and Section \ref{sec:conclusion} concludes the paper.

%% file: secs/background.tex
\section{Background}
\label{sec:background}

\subsection{Blockchain}

Blockchain was presented in 2008 with Bitcoin \cite{nakamoto2008bitcoin}. In short, block-chain is a decentralized storage system. Blockchain mainly consists of three layers: storage layer having data, network layer by which nodes connect to each other, and consensus layer used to keep the consistency.

In storage layer, data is split into transactions, a type of structured data. Transactions are managed by Merkle Trees, and packed into blocks. Each block hashes itself and id of the previous block selected to a hash value as its id. So blocks are linked by hash, and the chain of blocks appears. Because of these, the storage system is called blockchain. With new blocks appended, the chain grows. To modify or delete the data, namely transactions or blocks, is very difficult.

The network of blockchain is a peer-to-peer network, each node has a copy of data. Because of network delay, the network is usually asynchronous. Blockchains are roughly categorized into public blockchain, private blockchain and consortium blockchain. All cryptocurrency blockchains are pubic blockchains, in which all users can commit transactions and all records are public. A private blockchain is regarded as a centralized network since it is fully controlled by an organization. Recently, consortium blockchains develop rapidly to deal with transactions among several organizations, like Hyperledger fabric \cite{androulaki2018hyperledger}.

Because blocks can be just appended, blockchain uses consensus mechanism to keep the consistency. Each node can collect transactions sent by all nodes, pack them into a block and broadcast it to the network. The block will not be appended until most of nodes verify and save it to the storage. To encourage this kind of contribution, blockchain usually rewards the user whose node generates a block accepted by the network. The mechanism of generating and verifying blocks is called consensus mechanism, which depends on the consensus algorithm.

Besides, blockchain provide a script language to make transactions and be easily resolved. With it, more features can be achieved, like payment channel or muti-signature transaction. But the low readability affects its use.

In summary, the blockchain can be seen as a decentralized storage system with a FIFO writing mechanism. Based on mechanisms of storage, network and consensus, blockchain takes advantage of the immutability, auditability, persistency and decentralization, to provide safe, persistent and anonymous services.

\subsection{Smart Contract and Ethereum}

Because of the lack of functions of script language, developers need more powerful programming language and execution mechanism. A new concept, smart contract, was proposed \cite{szabo1994smart}. Smart contract was described as a set of digitally defined promises, including the process of performing them \cite{szabo1996smart}. In implementation, smart contract is a set of programs executed on blockchain, which receive requests, call pre-defined functions and save results in the blockchain.

Ethereum is the first blockchain providing Turing-complete programming language to develop smart contracts. Today, Solidity, a Javascript-like object-oriented programming language, is the most popular programming language in Ethereum community. A smart contract is developed by Solidity, compiled into bytecode, and then able to be deployed in the blockchain. To adopt compiled codes, Ethereum presents a new kind of users, contract user. A contract user has same features like individual user, but is controlled by the smart contract. Transactions to contract users will be converted to requests to contracts. All data of a contract user is open, including its transactions and bytecode of the smart contract.

Like transactions in blocks, each node of the blockchain contains a copy of a space for each user to save the data generated by smart contract execution, called storage. When a node receives a new block and does verification, it will check transactions one by one and convert transactions of contract users to contract requests then. In execution, the node loads request parameters and bytecode of the smart contract, executes the program and changes storage of the contract user. So smart contracts can run anywhere and the results are synchronized.

Because of Turing-complete, developers can implement endless programs in smart contracts, which cause that nodes executing them break down. To avoid it, Ethereum uses gas mechanism: each opcode in bytecode will cost a few Ethers (ETH, cryptocurrency that Ethereum issues), called gas, in execution. By the mechanism, Ethereum unifies contract requests and ETH transactions to transactions. Requester need to pay gas for the execution, but only the node generating the block that contains the request will get these Ethers. Deploying a smart contract is seen as a transaction and requires paying gas as well.

Thanks to these mechanisms, Ethereum succeed to implement smart contracts in the blockchain. Developers have ability to implement more features. Furthermore, a new kind of applications, decentralized application, appears in the network.

\subsection{Decentralized Application}

With growth of internet computing \cite{mei2012internetware, huang2013new}, especially blockchain and smart contracts, decentralized application becomes a new type of applications. They can have clients, mobile apps, servers like original applications, but the key data and operations are kept in smart contracts in a blockchain. Dapps interact with smart contracts by transactions, namely contract requests, and provide services based on them.

In theory, a single smart contract can be seen as a dapp. So it is difficult to find dapps by only smart contracts in the blockchain. Fortunately, there are some websites collecting dapps. Dapp authors usually submit their dapps to these websites to do advertisements, so we can get rough information about dapps from them. But there are not any websites can provide information of all of dapps. \textit{State of the ÐApps} \cite{stateofthedapps} is the most famous website tracking dapps.

%% file: secs/dataset.tex
\section{The Dataset}
\label{sec:dataset}

In this section, we briefly introduce main websites collecting dapps and smart contracts, and describe the features of our dataset collected from \textit{State of the ÐApps}, \textit{DappRadar} \cite{dappradar} and \textit{Etherscan} \cite{etherscan}. All data we collect are open and anonymous.

\subsection{Data Sources}

Our dataset consists of three parts: dapp information, related transactions and related smart contracts. We collect dapp information from \textit{State of the ÐApps} and \textit{DappRadar}, and get transactions and part of smart contract source codes from \textit{Etherscan}.

\subsubsection{\textit{State of the ÐApps} and \textit{DappRadar}}

\textit{State of the ÐApps} is a website that collects many decentralized applications. It is a privately funded and independent project. The project was founded in 2017 and has grown to be the main directory website of dapps and referenced by the homepage of Ethereum \cite{ethereumproject}. Dapp developers can submit their dapps to the website and get them published. It is free and provides many services for users and developers, such as filtering dapps and user discussions.

\begin{table}[!t]
    \caption{Top 10 Dapps in \textit{State of the ÐApps}}
    \label{tab:sodapp}
    \centering
    \begin{tabular}{lr}
        \hline
        {\bf \small Title} & {\bf \small Number of Transactions} \\
        \hline
        ForkDelta & 9,575,884 \\
        IDEX & 3,862,946 \\
        CryptoKitties & 3,130,573 \\
        Bitcoinereum & 1,451,752 \\
        OmiseGO & 1,271,880 \\
        Storj & 769,872 \\
        Ethereum Name Service & 456,702 \\
        Status & 407,946 \\
        PoWH 3D & 327,353 \\
        Ether Goo & 312,140 \\
        \hline
    \end{tabular}
\end{table}

For each dapp, the website provides it's title, description, categories, submit time and smart contract addresses. It summaries transactions daily and generates sparklines of users, transactions and volumes. The website marks dapps with categories and badges (e.g. status), and ranks them by dau (the number of daily active unique addresses). Table \ref{tab:sodapp} shows top 10 dapps on \textit{State of the ÐApps} and their numbers of related transactions from August 2017 to July 2018 respectively.

\textit{DappRadar} is another website of tracking dapps. It collects dapps by developers' submission as well. \textit{DappRadar} starts from November 22, 2017, and has collected more than 1,000 dapps and their information. It provides services like those of \textit{State of the ÐApps}, and all time charts of users and transaction volumes in addition. Almost all major dapps are submitted to it, but some new dapps are submitted to \textit{State of the ÐApps} first. We collect data from it as supplements.

\subsubsection{\textit{Etherscan}}
\label{sec:etherscan}

\textit{Etherscan} is a block explorer and analytic platform for Ethereum. It collects blocks, transactions, smart contract source codes and user addresses from Ethereum blockchain, and display them in a web UI. \textit{Etherscan} starts at 2015, is based on Ethereum blockchain, and integrates proposals and standards from communities, and now provides lots of useful services, including apis, charts and smart contract development tools. We can get source codes and related transactions by user addresses and hash of transactions.

Smart contracts providing source codes is named verified contracts, the codes are submitting by their developers. There are still lots of contracts without source codes, we will get empty return values when calling apis.

\subsection{Data Collection}

\begin{table}[!t]
    \caption{Dataset}
    \label{tab:dataset}
    \centering
    \begin{tabular}{lr}
        \hline
        {\bf \small Number of dapps} & 1,454 \\
        {\bf \small Number of dapps providing contracts} & 734 \\
        {\bf \small Number of contracts} & 2,740 \\
        {\bf \small Number of open-source contracts} & 1,597 \\
        {\bf \small Number of related transactions} & 25,954,362 \\
        {\bf \small Number of users' addresses} & 1,271,766 \\
        {\bf \small Volume of all transactions (ETH)} & 5,953,683.954 \\
        \hline
    \end{tabular}
\end{table}

Until September 2018, there are over 1,500 dapps on \textit{State of the ÐApps}. Each dapp is associated with some smart contracts and transactions. We collect dapps submitted before August 1, 2018, and one year usage data from August 1, 2017 to July 31, 2018. Table \ref{tab:dataset} shows meta data of our dataset.

Dapp data consists of two types: (1) dapp information (i.e., name, authors, categories, addresses of smart contracts and so on) (2) smart contract information (i.e., creator address, source codes and so on). Usage data is represented as transactions on Ethereum. 

\subsubsection{Dapp Information}
Dapp information comes from authors when submitting, including basic information, categories, contracts and social information. It is mainly from \textit{State of the ÐApps}. In \textit{State of the ÐApps}, dapps is roughly divided into some categories. Almost all dapps belong to a category, but some one has more than one or no categories. Dapps are ranked by dau. Social information is contact information of some social websites such as \textit{Github}, \textit{Twitter} and \textit{Facebook}.

Dapps from \textit{DappRadar} is a subset of those from \textit{State of the ÐApps}, so we collect data from it as supplements.

Finally, we select 1,454 dapps having information from \textit{State of the ÐApps} and \textit{DappRadar}. Only 734 dapps provide addresses of smart contracts. In our dataset, we rank them by the numbers of related transactions in a year rather than dau that websites usually use.

\subsubsection{Smart Contract Information}

For developers, Ethereum provides solidity, a JavaScript-like programming language. Almost all smart contracts are developed with it. From dapp information we can get addresses of smart contracts, by which smart contract information can be collected from \textit{Etherscan}, including source code written by solidity.

Smart contracts don't occupy equal status in a dapp: they are marked with mainnet, additional and so on. But we see them as the same, because in the study of this paper it's difficult and unnecessary to differentiate them.

In dapp information data, we collect 2,740 unique smart contract addresses and get their source codes. To get in-depth results, we extract md5. Because the source codes are not formatted, we manually split codes and count the lines that are not empty as LoC.

\subsubsection{Transaction Data}

Both \textit{State of the ÐApps} and \textit{DappRadar} provide summary of related transactions, but neither are incomplete: \textit{State of the ÐApps} just provides monthly sparklines and quarterly statistical values; \textit{DappRadar} provides sparklines just from November 22, 2017 of users and volumes. So we need to get all transactions from August 1, 2017 to July 31, 2018. Apis of \textit{Etherscan} allow us to get related transaction data by contract addresses. Data of a transaction consists of addresses of two transactors, time stamp, input of a contract call, gas and value. It is the only usage data we can collect.

In summary, we get 25,954,362 transactions by smart contract addresses, and 1,271,766 unique user addresses.

%% file: secs/descriptive.tex
\section{A Descriptive Analysis}
\label{sec:descriptive}

In this section, we make a descriptive analysis on dapp distribution. In websites, dapps are usually ranked by their dau. We use four metrics to make more comprehensive analysis: (1) category; (2) submit time of dapp; (3) the number of unique addresses (users); (4) the number of transactions; (5) transaction volume. The former two metrics can indicate how dapps are designed and submitted, and the latter three can indicate how much a dapp is really used.

\subsection{Popularity by Categories}

\begin{table}[!t]
    \caption{Categories and Their Summaries of Transactions (Sorted by Transactions)}
    \label{tab:categories}
    \centering
    \begin{tabular}{lrrr}
        \hline
        {\bf \small Category} & {\bf \small Dapps} & {\bf \small Unique Users} & {\bf \small Transactions} \\
        \hline
        Exchanges & 126 & 613,258 & 15,519,946 \\
        Wallet & 18 & 684,880 & 11,479,517 \\
        Games & 350 & 145,268 & 4,953,632 \\
        Finance & 95 & 400,480 & 1,709,780 \\
        Storage & 148 & 86,938 & 874,628 \\
        Security & 89 & 75,303 & 770,413 \\
        Gambling & 19 & 46,064 & 720,165 \\
        Social & 21 & 159,740 & 643,895 \\
        Identity & 9 & 17,725 & 489,724 \\
        Development & 11 & 30,803 & 243,706 \\
        Governance & 53 & 53,175 & 197,428 \\
        Media & 19 & 5,786 & 42,400 \\
        Property & 23 & 2,911 & 41,094 \\
        Health & 2 & 8 & 38 \\
        Energy & 3 & 20 & 35 \\
        Insurance & 1 & 0 & 0 \\
        Uncategorized & 58 & 58 & 1,891,574 \\
        All dapps & 734 & 1,271,766 & 25,954,362 \\
        \hline
    \end{tabular}
\end{table}

First, we investigate categories of dapps. \textit{State of the ÐApps} divide dapps into 16 categories. Their summaries of transactions are shown in Table \ref{tab:categories}. Because there are some dapps having multiple categories and some having none, summary of all dapps is not equal to sum of all categories.

\begin{figure}[!t]
    \subfloat[Dapp Distribution]{
        \includegraphics[width=0.22\textwidth]{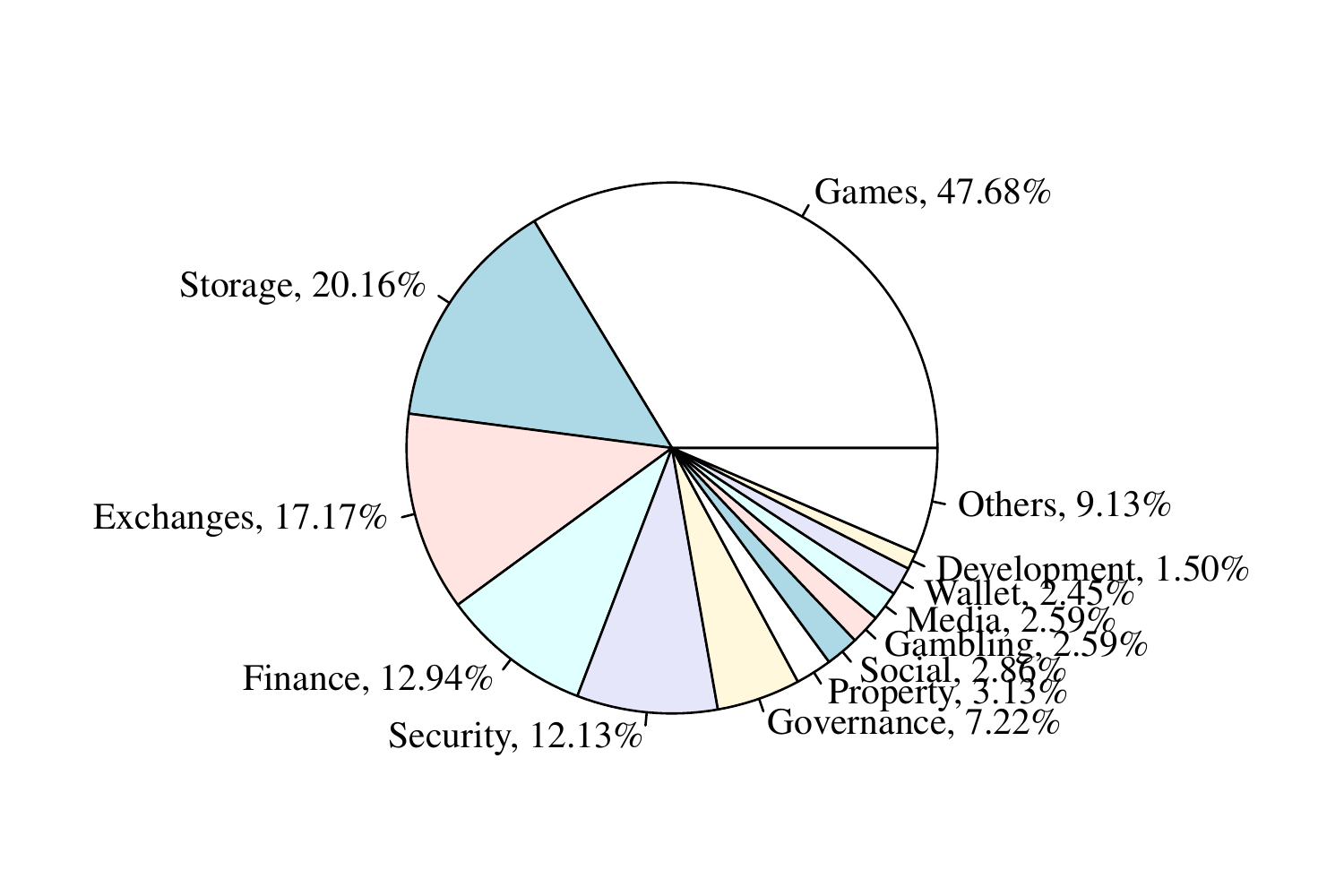}
        \label{fig:category_pie_dapp}
    }
    \subfloat[User Distribution]{
        \includegraphics[width=0.22\textwidth]{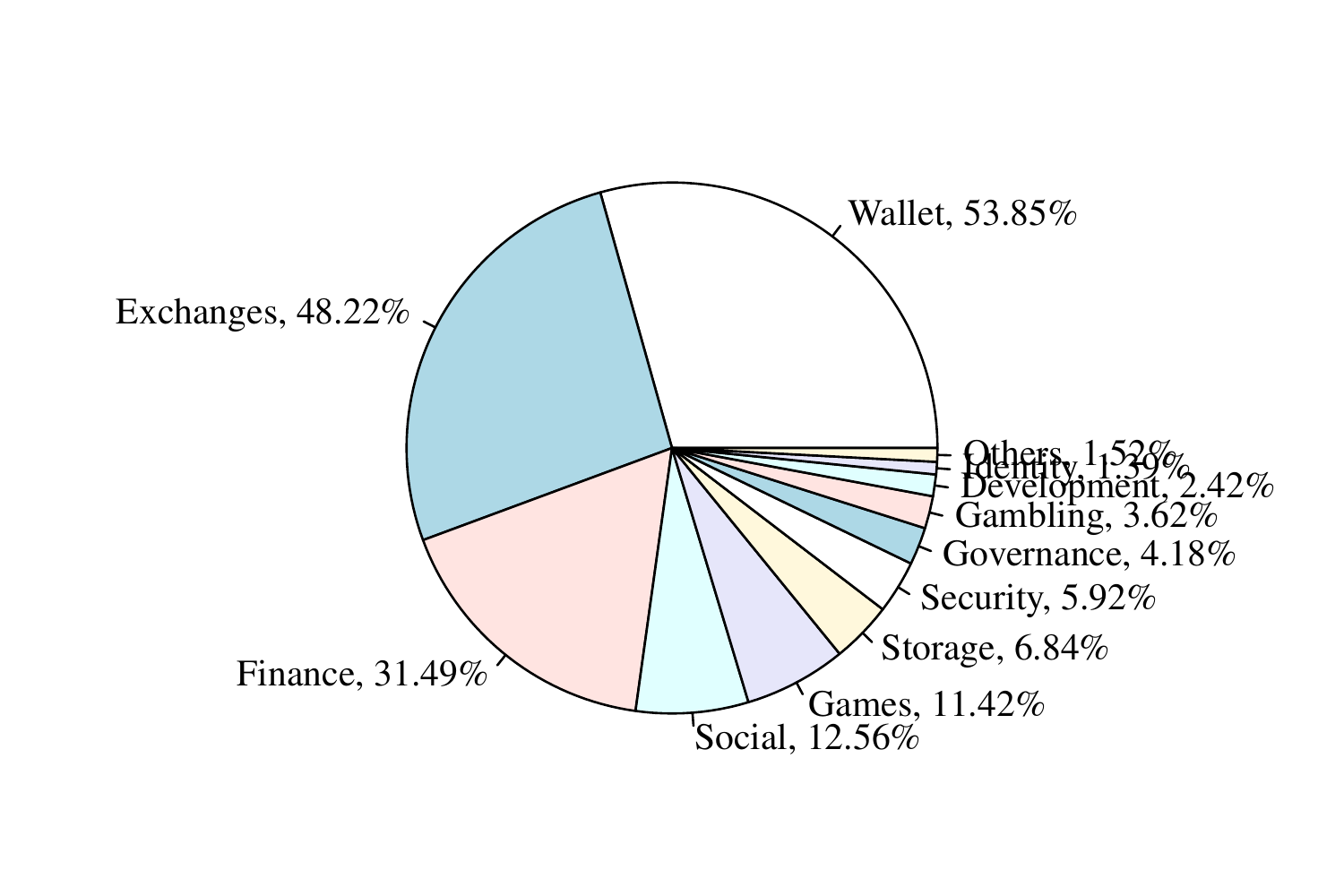}
        \label{fig:category_pie_user}
    }
    \hfill
    \subfloat[Transaction Distribution]{
        \includegraphics[width=0.22\textwidth]{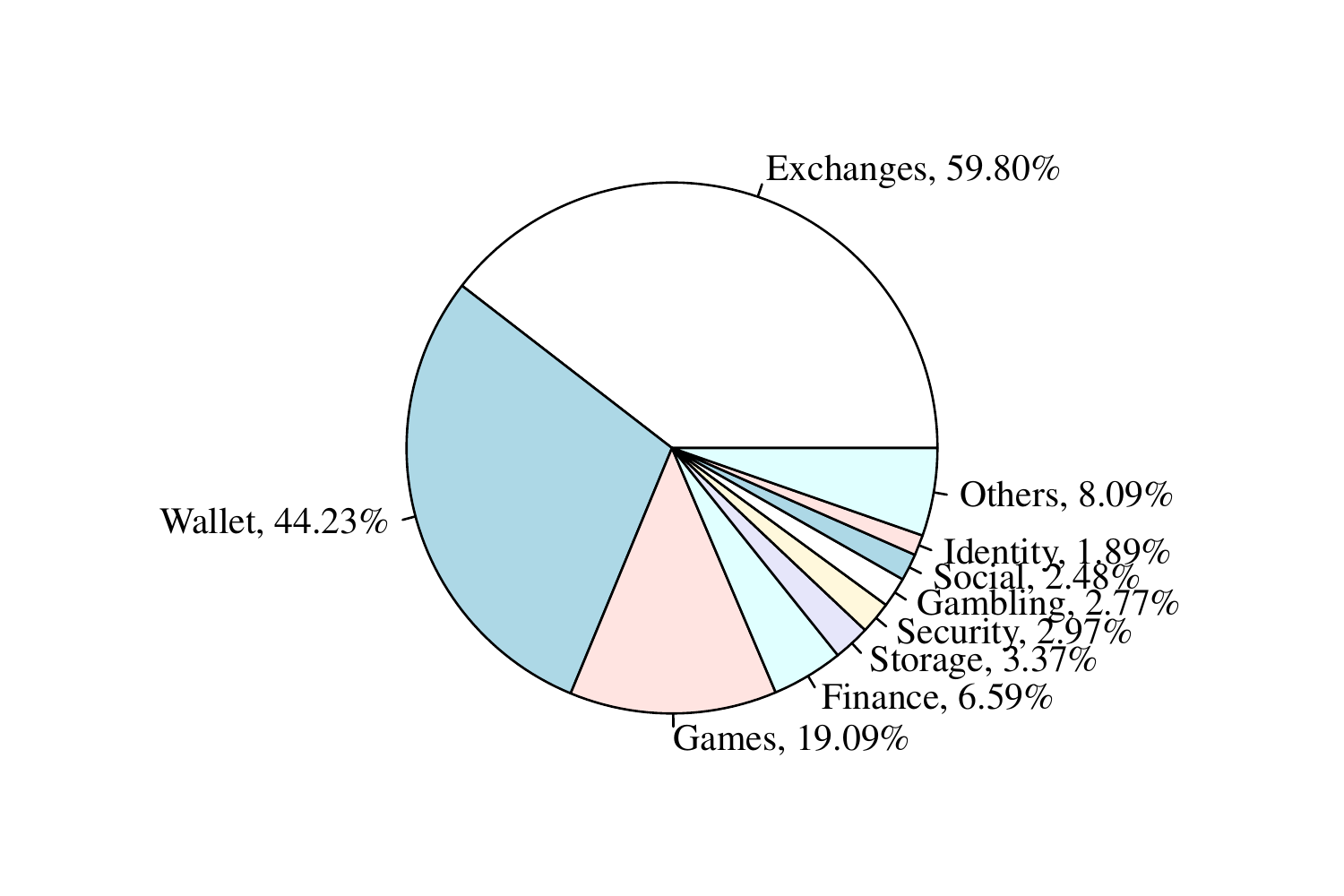}
        \label{fig:category_pie_txn}
    }
    \subfloat[Transaction Volume Distribution]{
        \includegraphics[width=0.22\textwidth]{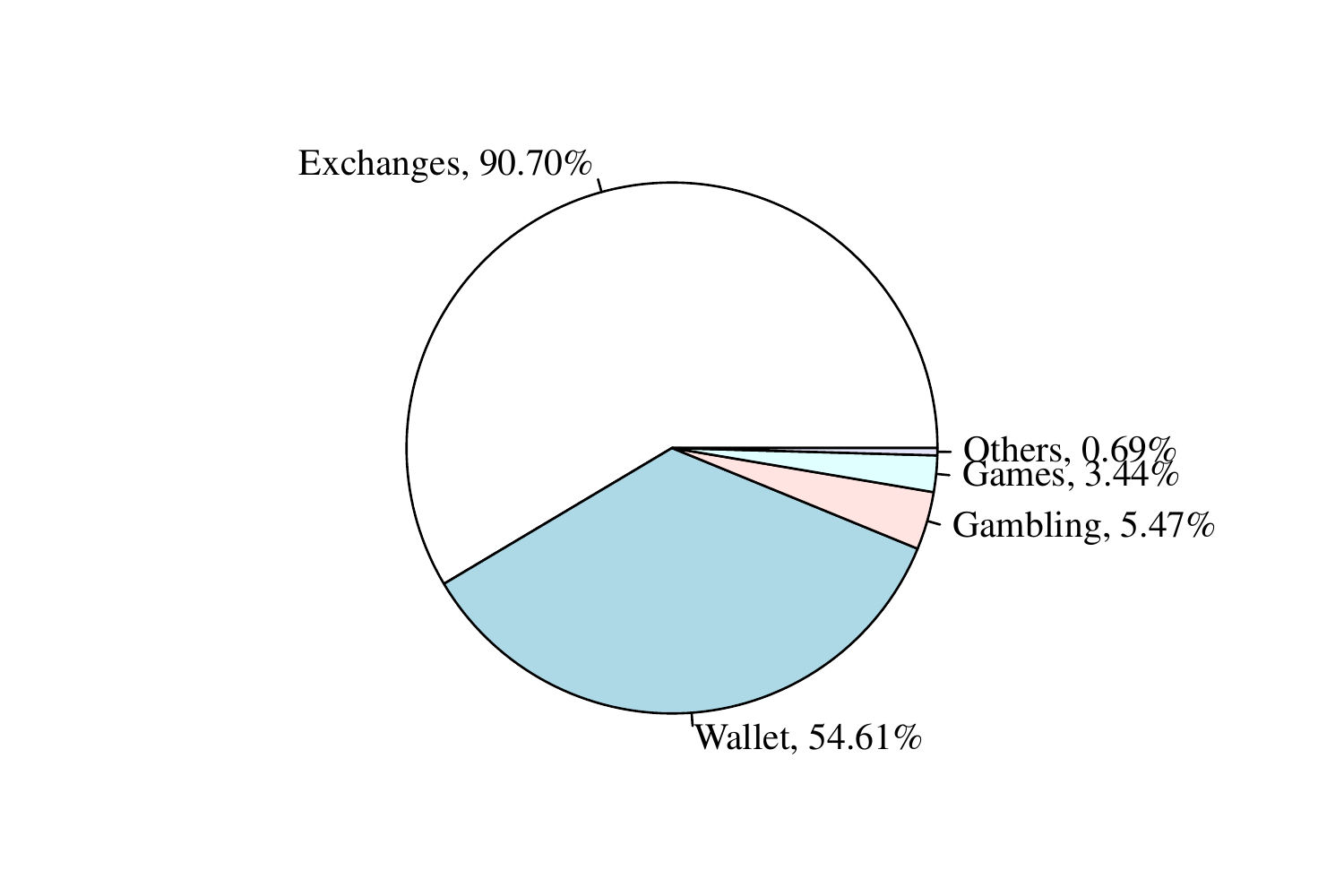}
        \label{fig:category_pie_vol}
    }
    \caption{DAPP Popularity by Category}
    \label{fig:category_pie}
\end{figure}

Figure \ref{fig:category_pie} shows that category Wallet has 2.45\% dapps and the most users, as well as category Exchanges has 17.17\% dapps and the most transactions and transaction volume. Because of financial characters of Ethereum, dapps of Wallet and Exchanges are published most and have great influence. Many cryptocurrencies are issued through Ethereum smart contracts, Exchanges dapps help users to transfer them and Wallet dapps are used to manage them. 47.68\% of dapps belong to category Games and have 11.42\% users, 19.09\% transactions and 3.44\% transaction volume. In top 10 dapps in Table \ref{tab:sodapp}, three of them are of category Wallet, 2 are of category Exchanges, and one of them has the two categories.

Category Health, Energy, Insurance have small percentage as they are newer categories, and represent the direction of dapp development.

\subsection{Growth over Time}

\begin{figure}[!t]
    \centering
    \includegraphics[width=2.5in]{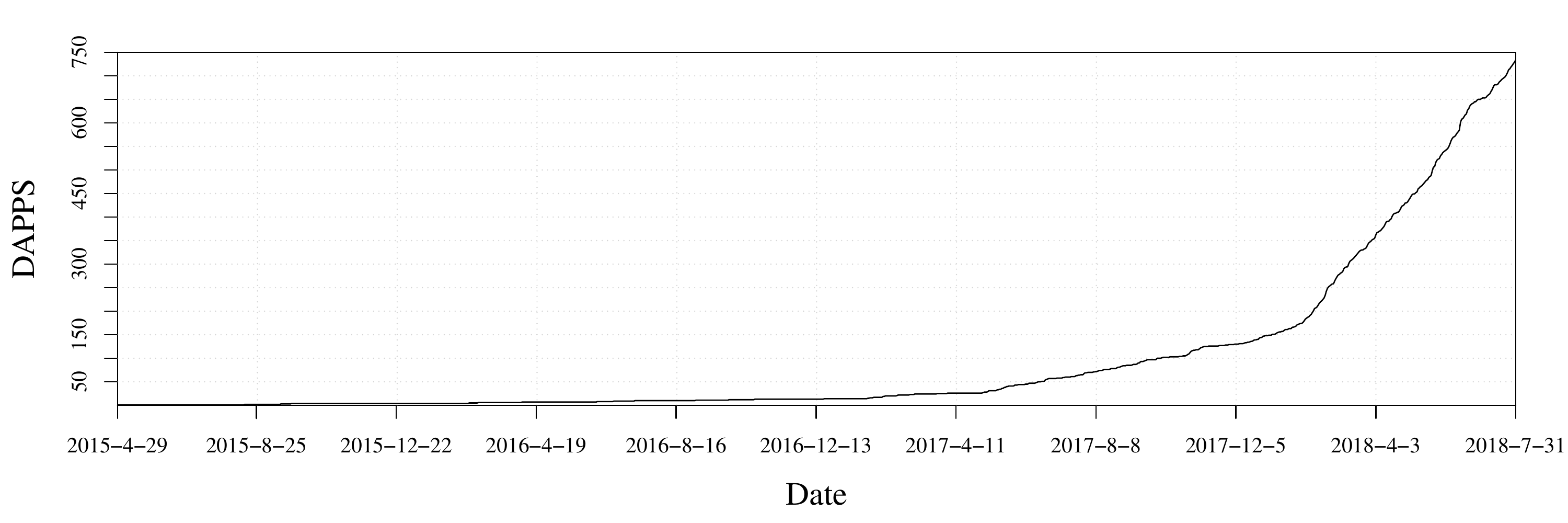}
    \caption{Dapp's Growth over Time}
    \label{fig:all_dapp_grow}
\end{figure}

Figure \ref{fig:all_dapp_grow} shows how the number of dapps grows over time. In our dataset, the first dapp is submitted at April 29, 2015. Dapps grows significantly from May, 2017. After January, 2018, more dapps are submitted every day. In June 14, 2018, the number of submitted dapps is 17, it is the max before July 31, 2018. Compared to ETH price curve, its trend is similar.

\begin{figure}[!t]
	\centering
    \subfloat[]{
        \includegraphics[width=2.5in]{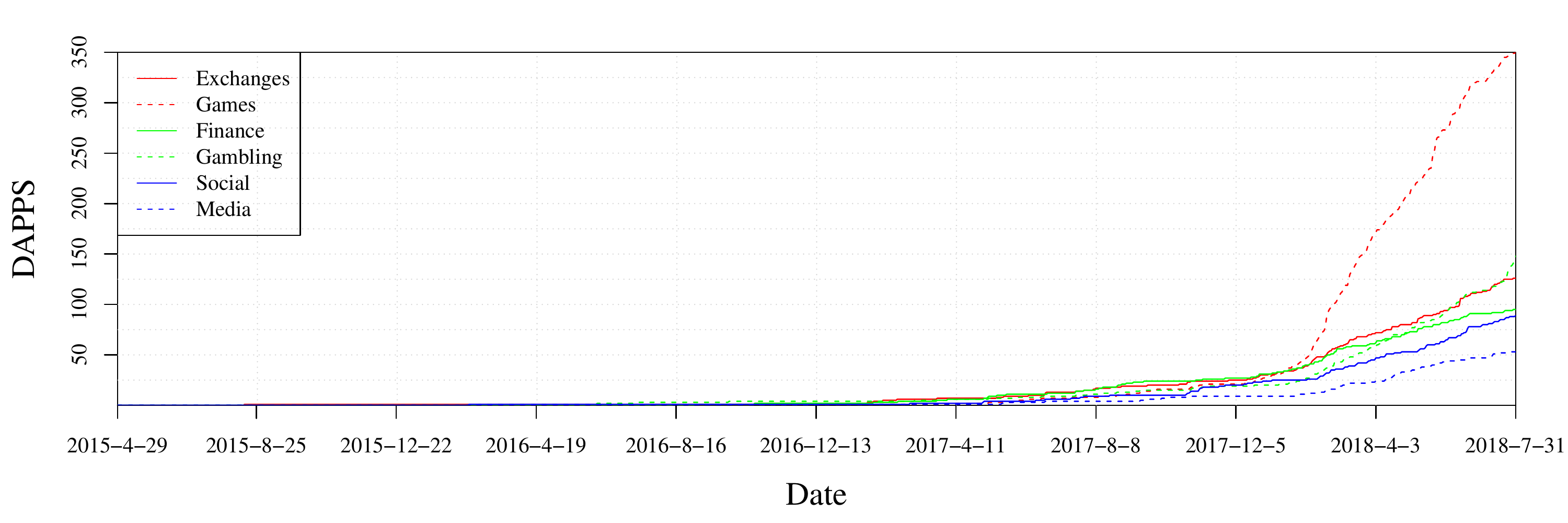}
        \label{fig:category_grow1}
    }
    \hfill
    \subfloat[]{
        \includegraphics[width=2.5in]{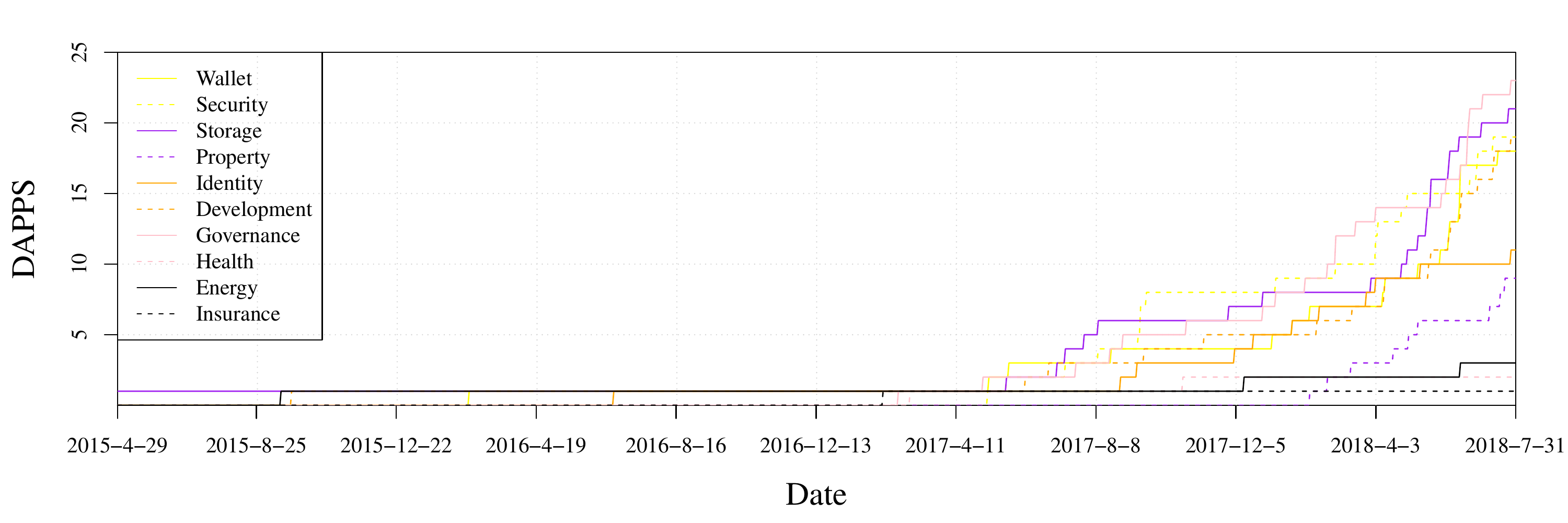}
        \label{fig:category_grow2}
    }
	\caption{DAPP's Growth over Time by Category}
	\label{fig:category_grows}
\end{figure}

Growth of all categories of dapps is shown in Figure \ref{fig:category_grows}. DigixGlobal of category Storage was the first submitted dapp in 734 dapps. At August 15, 2017, the second one, Augur was submitted and belongs to category Exchanges. The first dapp of category Wallet was submitted February 22, 2016.

At the beginning, all categories had similar trends. After May 16, 2017, the numbers of dapps of category Exchanges, Games, Gambling, Social, Finance grew rapidly. At January 15, 2018, category Games had the same number with category Exchanges and Finance, and then grew over two others. 
Now category Gambling is hot as well as category Exchanges. Growth of the number of dapps of category Finance slowed down and reachs a similar level as category Social. There were not new dapps of category Media submitted from October 25, 2017 to January 25, 2018, then the number of them slowly grows over 50.

About other categories, their first dapps were submitted before May 9, 2017 except category Property. But their numbers of dapps grew slowly before May, 2017. After February 6, 2018, all categories had at least one dapp, but at July 31, 2018, category Health, Energy, Insurance have less than 4 dapps.

\subsection{Popularity by Users}


\begin{figure}[!t]
	\centering
    \subfloat[Percentage of Users against Dapp Rank]{
        \includegraphics[width=0.22\textwidth]{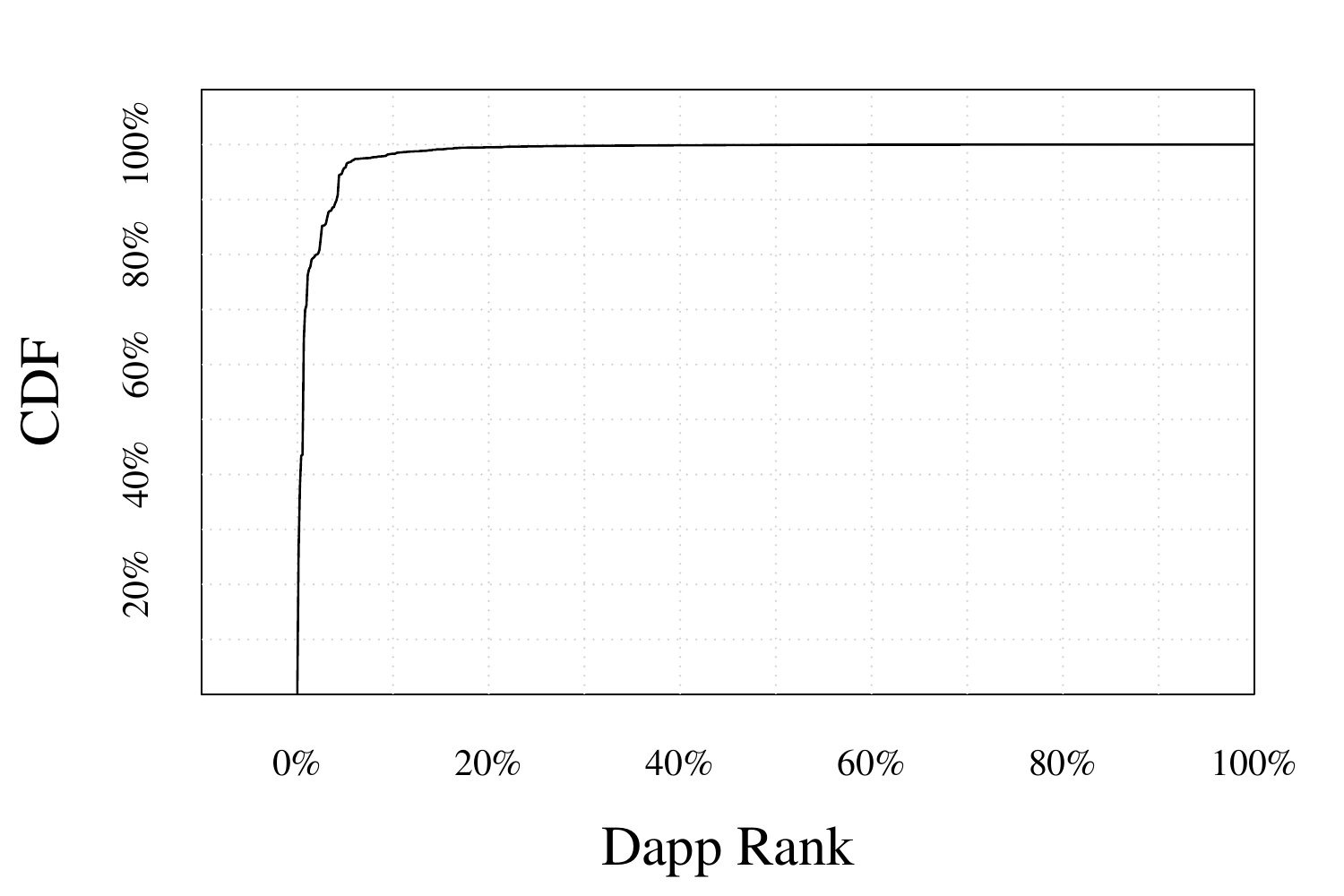}
        \label{fig:rank_users}
    }
    \subfloat[Users of a Dapp]{
        \includegraphics[width=0.22\textwidth]{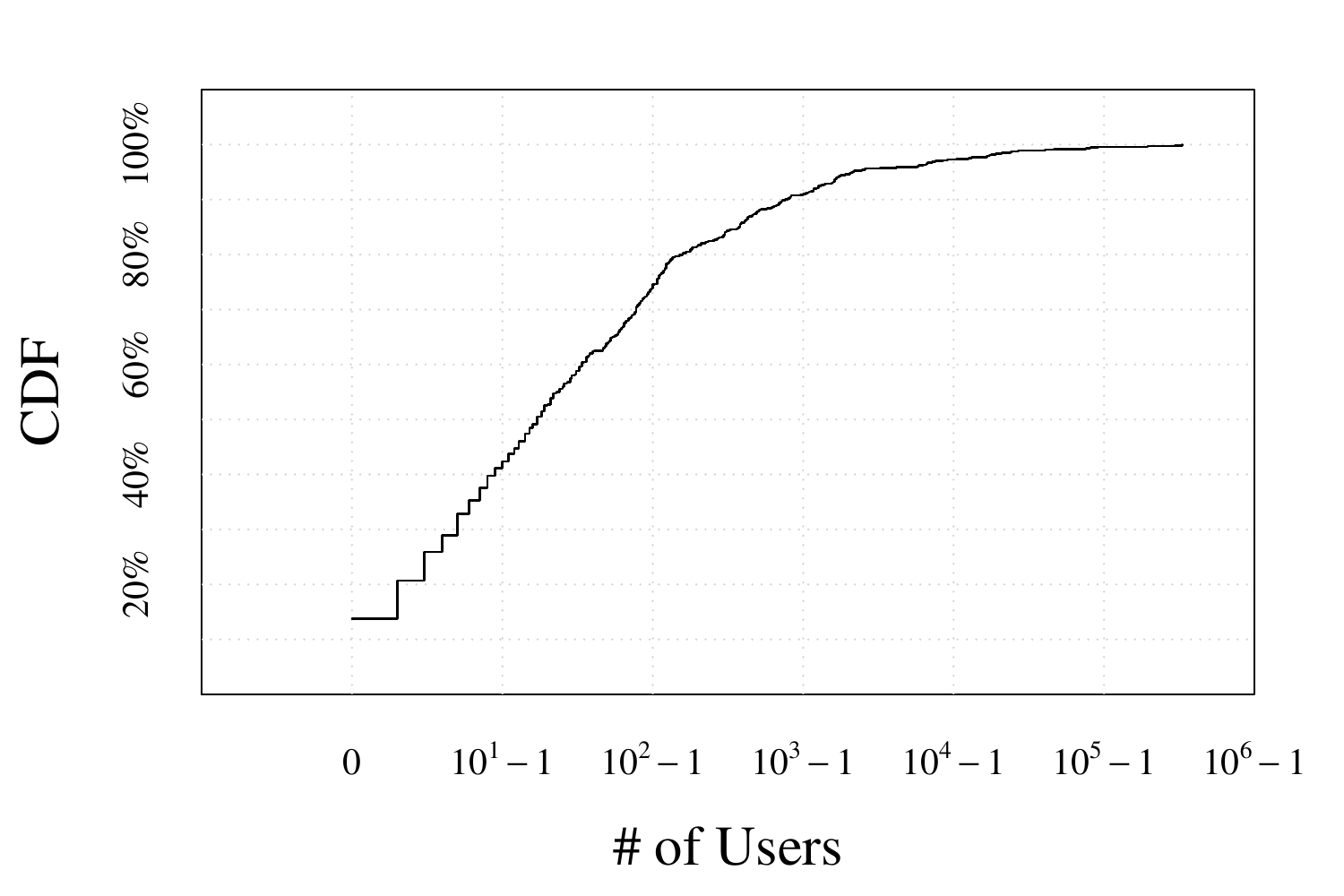}
        \label{fig:cdf_users}
    }
	\caption{DAPP Popularity by users}
	\label{fig:popularity_users}
\end{figure}

We have investigated two metrics, category and submit time of the dapp. The results partly show how dapps are designed and submitted. Now we have to know how much a dapp is really used. In dapps, smart contracts are used to interact with blockchain and do core operations, such as transferring ETH and tokens (cryptocurrencies) and store data. So we judge whether a dapp is really used, by transactions in Ethereum blockchain. We collect related transactions of dapps' smart contracts, extract "from" addresses or "to" addresses that are not equal to related smart contract addresses as dapps' users. Because we can't relate a user address to a real man, we see a unique address as a user.

Figure \ref{fig:rank_users} shows that the cumulative distribution function (CDF) of the percentage of dapp users against the dapp ranking by transactions. The results show that the dapp users follow the Pareto principle, i.e. less than 20\% dapps have almost all users. We can conclude that the more a dapp is used, the more users it has.

We also explore the distribution of users of each dapp. Figure \ref{fig:cdf_users} indicates that about 75\% dapps are used by only less than 100 users.

There are some dapps have relatively few users but many transactions, such as Bitcoinereum and KIWI (Bitcoin-like tokens). The two above have few transaction volumes as well. Users of these dapps may be cryptocurrency investors. Although such kind of dapps is rare, their transactions consume some traffic and affect other dapps.

\subsection{Popularity by Transactions}

\begin{figure}[!t]
	\centering
    \subfloat[Percentage of Transactions against Dapp Rank]{
        \includegraphics[width=0.22\textwidth]{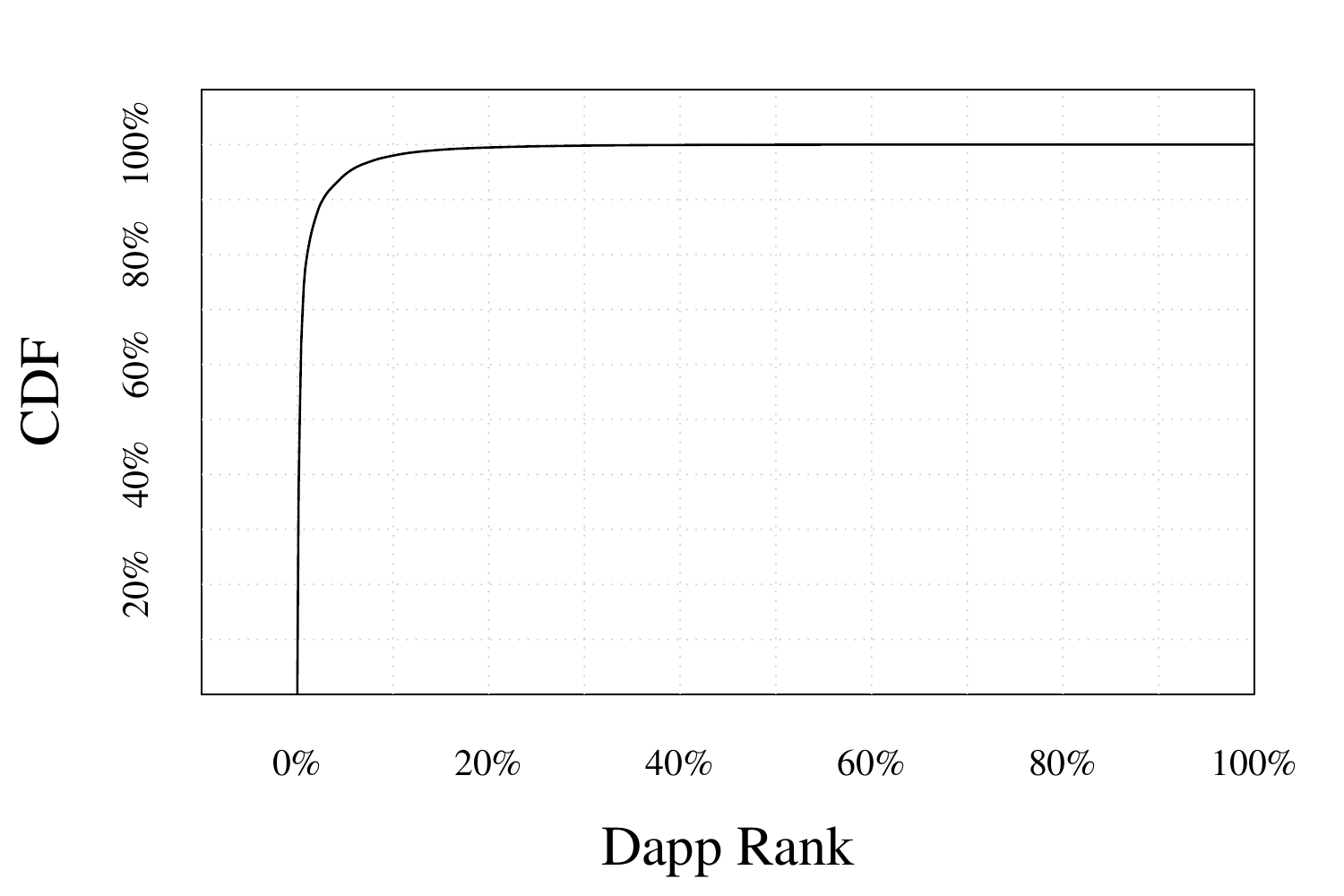}
        \label{fig:rank_trans}
    }
    \subfloat[Transactions of a Dapp]{
        \includegraphics[width=0.22\textwidth]{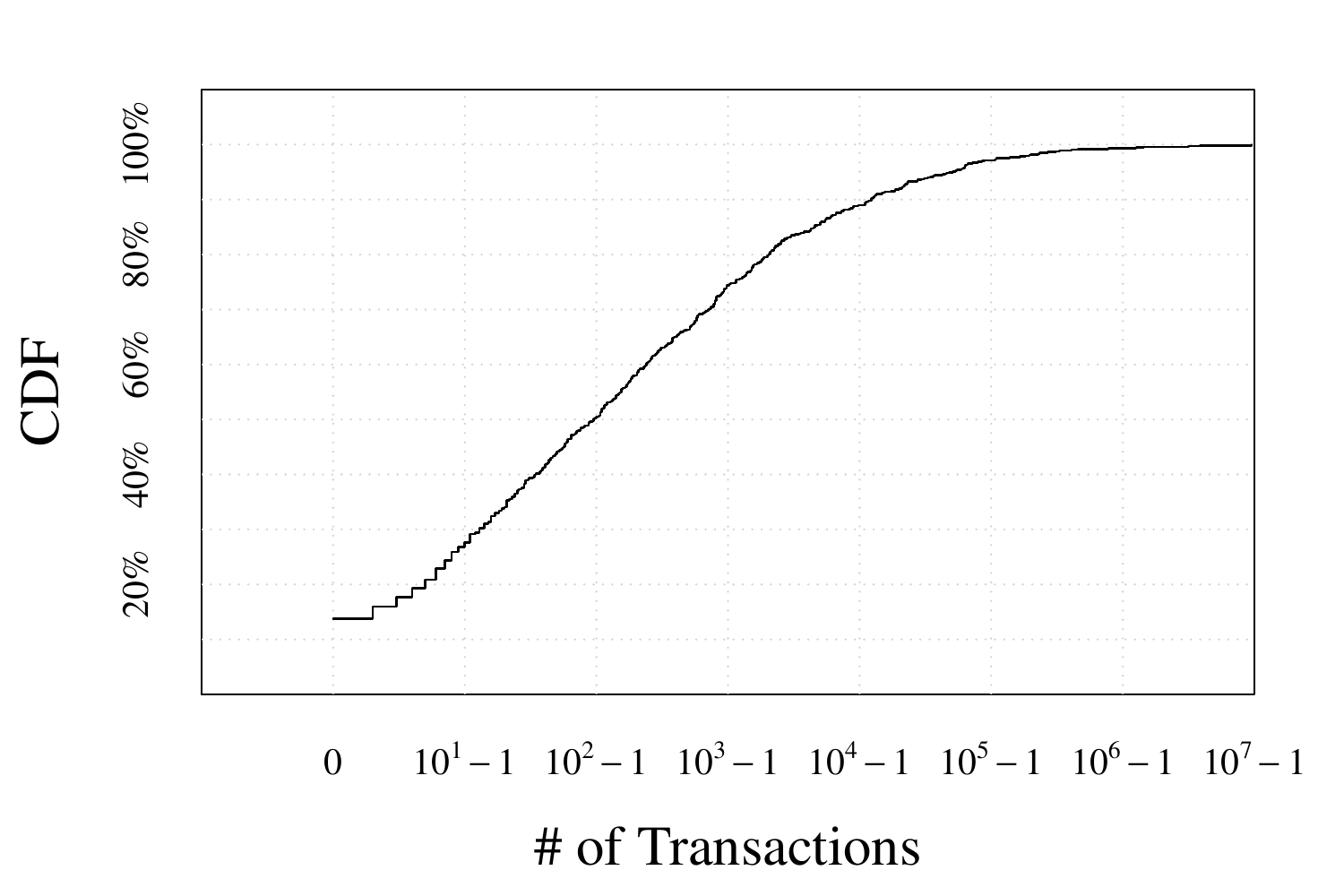}
        \label{fig:cdf_trans}
    }
	\caption{DAPP Popularity by transactions}
	\label{fig:popularity_trans}
\end{figure}

Because all transactions are added to the block with paying some ETH, transactions can be seen as real behaviours of users. Figure \ref{fig:rank_trans} shows that the CDF of the percentage of transactions of dapps against dapp ranking by transactions. We can find that a few dapps have 80\% transactions, and almost all dapps have few transactions.
Figure \ref{fig:cdf_trans} shows that over 70\% dapps have less than 1,000 transactions. Such findings indicates a "long-tail" of dapps that are rarely active. Only 40 dapps of them were submitted before August 1, 2017, so dapps' submit time influences their popularity.

\subsection{Popularity by Transaction Volumes}

\begin{figure}[!t]
	\centering
    \subfloat[Percentage of Transaction Volumes against Dapp Rank]{
        \includegraphics[width=0.22\textwidth]{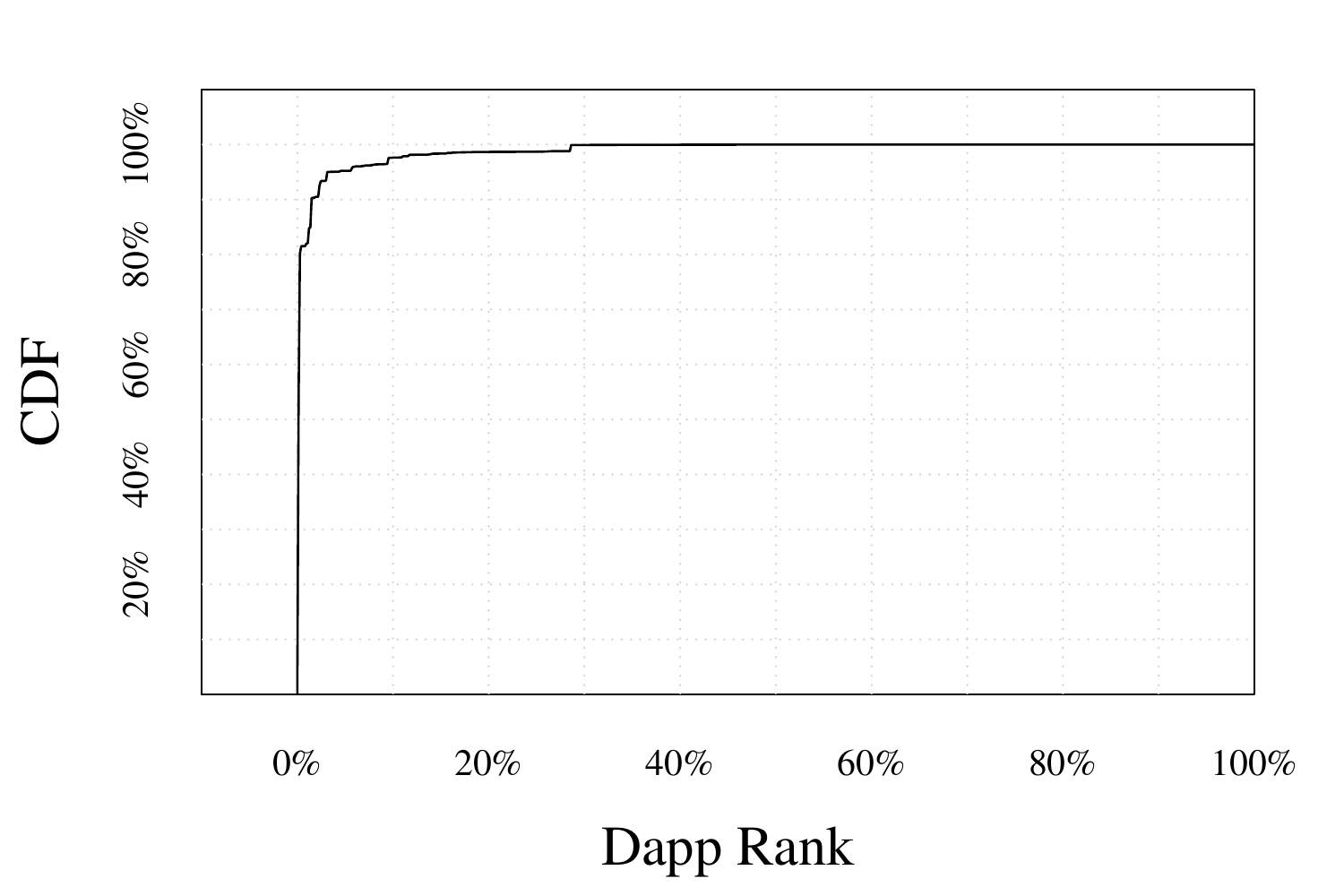}
        \label{fig:rank_volume}
    }
    \subfloat[Transaction Volume of a Dapp]{
        \includegraphics[width=0.22\textwidth]{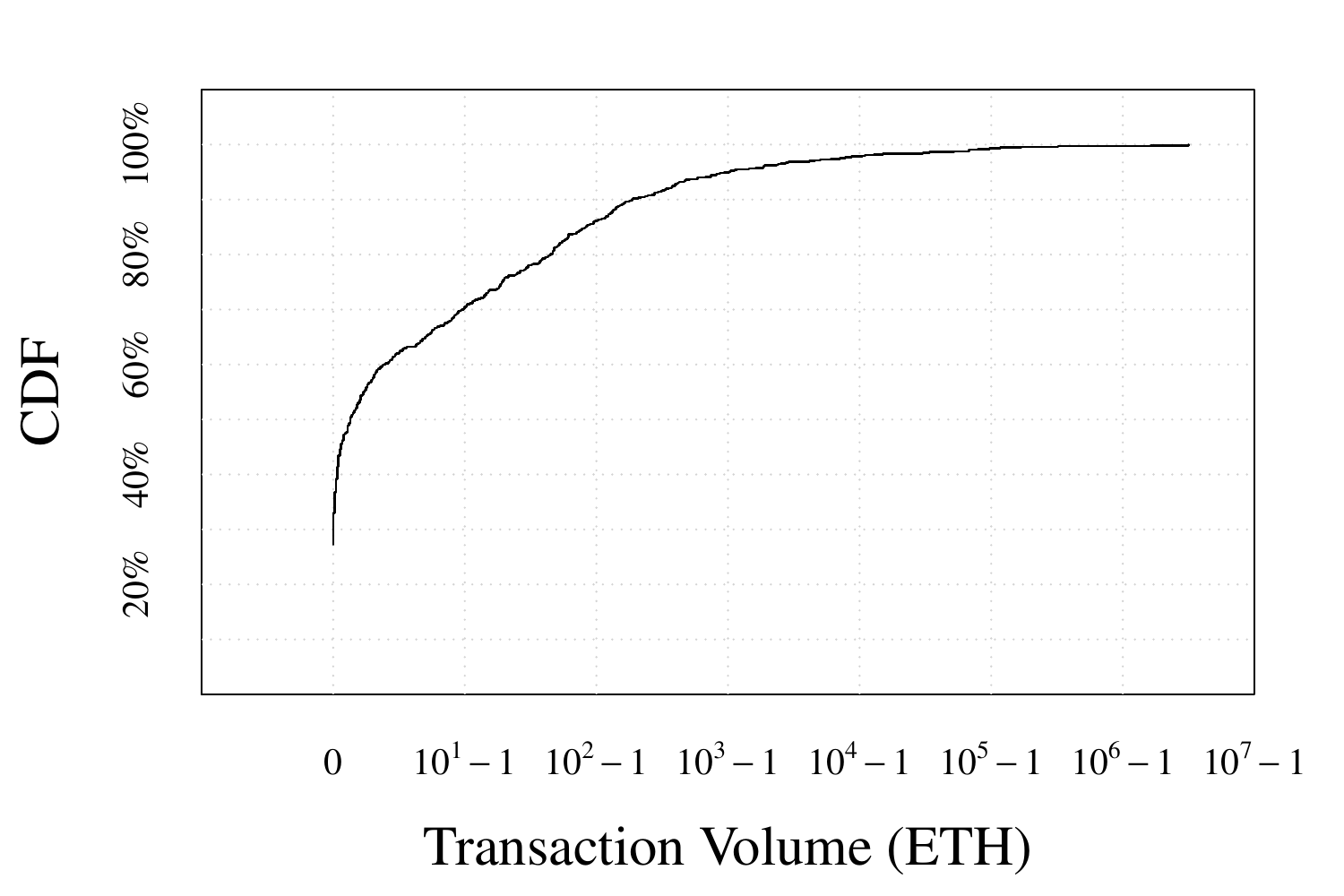}
        \label{fig:cdf_volume}
    }
	\caption{DAPP Popularity by ETH Trade}
	\label{fig:popularity_volume}
\end{figure}

In Ethereum, each transaction needs to pay some ETH to miners, but they can transfer 0 ETH from a user address to another. Therefore, dapps having many transactions may not be economic beneficial. Figure \ref{fig:rank_volume} illustrate the CDF of the percentage of dapp's transaction volumes against dapp rank. We can find that the Pareto principle also holds for dapp transaction volumes. And they are more concentrated: compared to Figure \ref{fig:cdf_users} and \ref{fig:cdf_trans}, about 5\% dapps have 94.65\% users, 94\% transactions and 95.22\% of transaction volume.

Figure \ref{fig:cdf_volume} shows that about 70\% dapps have only less than 10 ETH a year and 27.25\% dapps don't have transaction volume. 

%% file: secs/smartcontract.tex
\section{An In-Depth Analysis}
\label{sec:sc}

In this section, we investigate how dapps are developed, e.g., whether the dapp is usually open-sourced, how many contracts a dapp has, how many lines in a contract source code, how smart contract source codes are reused, and how to transfer a web application to a dapp.

\subsection{Open Source}

In general, source codes of dapps can be divided into two parts: smart contracts and the others. After development, smart contracts are compiled to bytecodes and then deployed to blockchain. Dapps provide end-users with services through UIs, do some operations and interact with blockchain through smart contracts. So open sources of dapps can be divided into two parts as well.
Any smart contracts can be gotten from blockchain straightly, but they are only bytecodes. Developers can submit source codes to block explorers or open source code repositories to make smart contracts open-source. Other parts of dapps can be open-source like general softwares, android apps or web apps.

\begin{figure}[!t]
    \centering
    \includegraphics[width=2.5in]{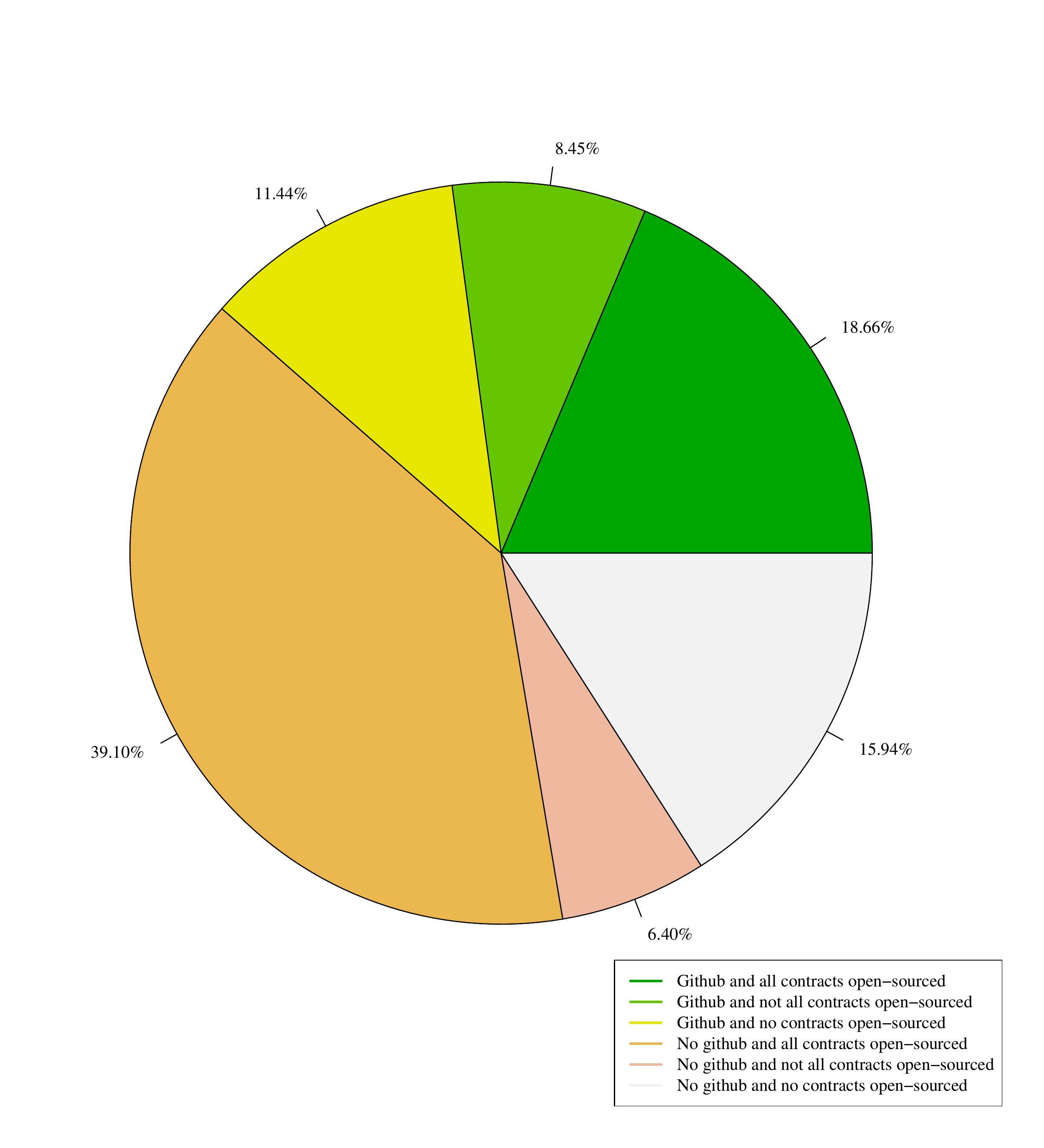}
    \caption{Percentage of Open-sourced Dapps}
    \label{fig:open_source}
\end{figure}

In our dataset, all open-sourced dapps are open-sourced through \textit{Github} and \textit{Etherscan}. Because bytecodes are not readable enough, developers usually submit source codes to improve the readability to gain trust of users. In figure \ref{fig:open_source}, we can find that 57.77\% of dapps providing contract addresses open their smart contract source codes. Considering that some smart contracts are reused and the source codes aren't updated immediately, 72.62\% of dapps may open their smart contract source codes. Only 18.66\% of dapps are fully open-sourced, and 15.94\% of dapps are not open-sourced.
61.44\% of dapps don't open their other codes, it means they hide their business related codes for users.

\begin{figure}[!t]
    \centering
    \includegraphics[width=2.5in]{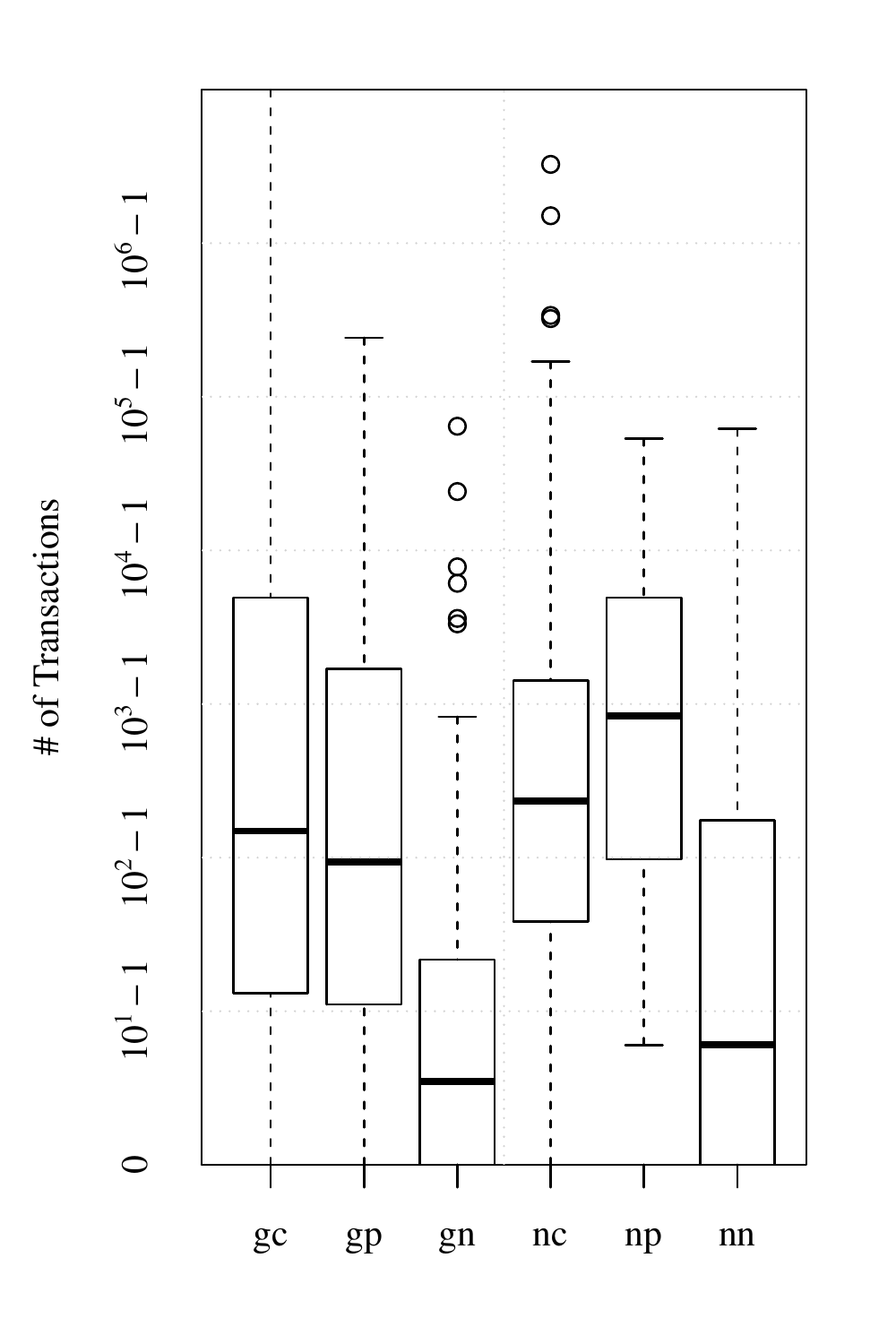}
    \caption{How does open source influence popularity of dapps}
    \label{fig:open_src_txn}
\end{figure}

We analyze relationship between open source and popularity of dapps further and get Figure \ref{fig:open_src_txn}. The left three boxes represent transactions of dapps open-sourced through \textit{Github} and right three represent those of dapps having not open-sourced themselves. In each three boxes, from left to right, boxes represent respectively all smart contracts open-sourced, smart contracts partly open-sourced and no smart contracts open-sourced. We can find that dapps whose smart contracts are open-sourced or partly open-sourced have lower median of transactions. Dapps with fully open-sourced smart contracts have higher maximum transactions.

So we can conclude that open source of smart contracts can improve dapps' popularity, and open source of business related codes has no significant influence. But in top 10\% dapps that have 98.33\% unique users, 97.97\% transactions and 97.62\% of transaction volume, 39.73\% of them are fully open-sourced.

\subsection{Distribution by Smart Contracts}

\begin{figure}[!t]
	\centering
    \subfloat[Percentage of Numbers of Smart Contracts against Dapp Rank]{
        \includegraphics[width=2.5in]{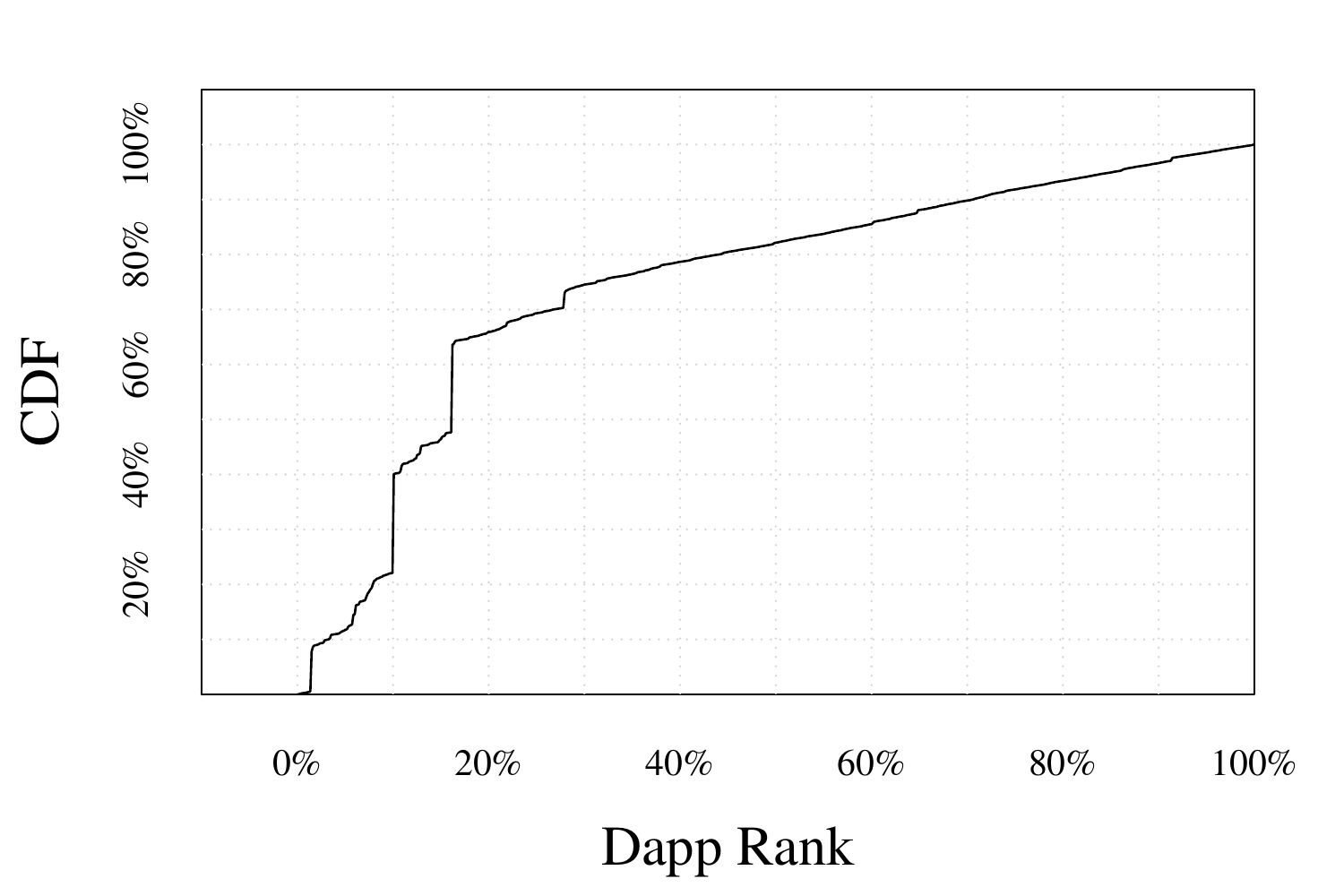}
        \label{fig:rank_sc}
    }
    \hfill
    \subfloat[The Number of Smart Contracts of a Dapp]{
        \includegraphics[width=2.5in]{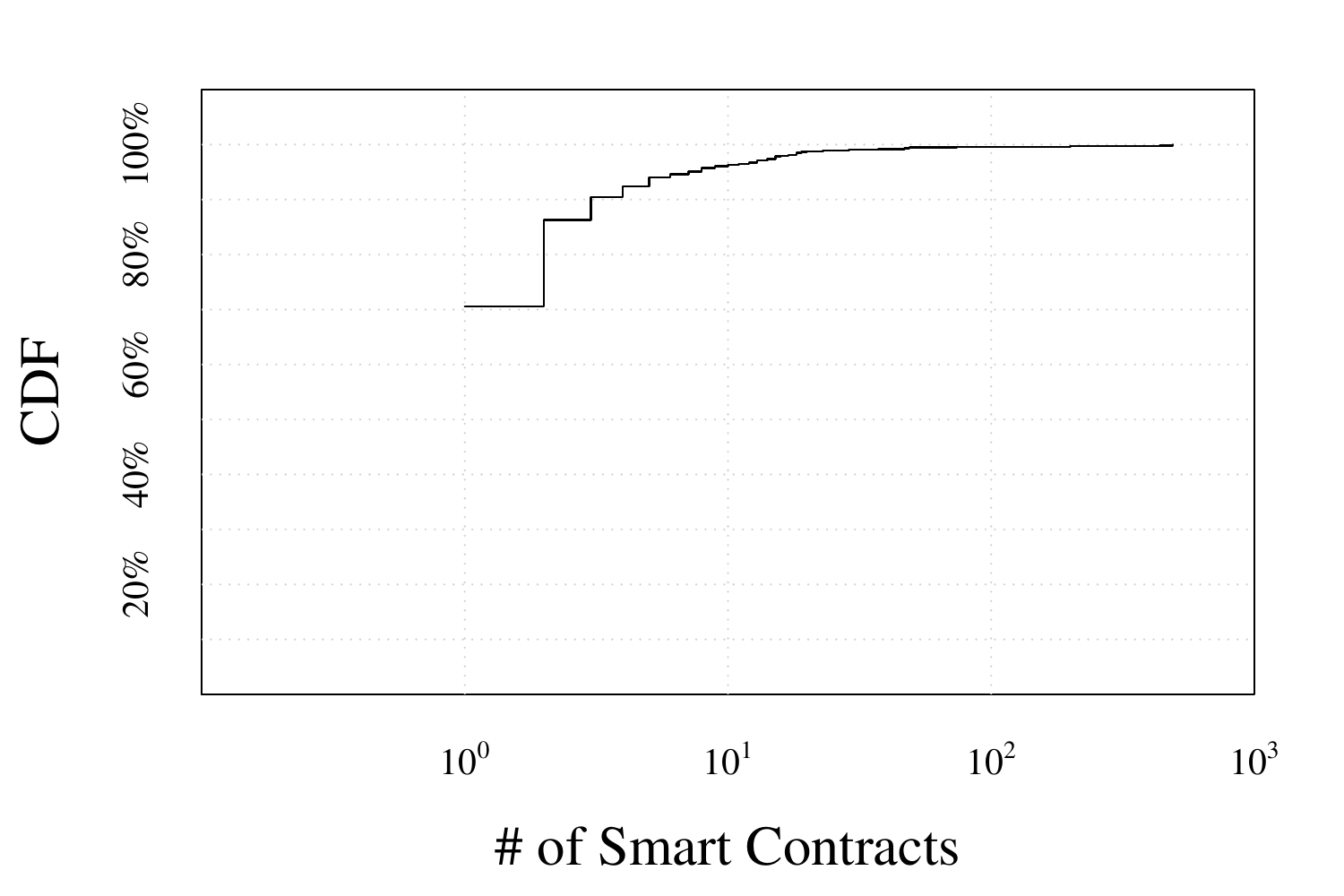}
        \label{fig:cdf_sc}
    }
	\caption{DAPP Popularity by Numbers of Smart Contracts}
	\label{fig:popularity_sc}
\end{figure}

Smart contracts work as core data storage and core operation handlers for dapps. To ensure that smart contracts work correctly and don't burden blockchain, Ethereum limits length and functions of smart contracts. Dapp developers sometimes have to divide features carefully to multiple smart contracts, especially data consist of low related several parts. So, the number of smart contracts of a dapp can indicate the complexity of functionality and the difficulty of development.

Figure \ref{fig:rank_sc} shows that the CDF of the percentage of smart contracts of dapps against the dapp rank. We can find that top 10\% of dapps have 22.08\% smart contracts and it is influenced by few dapps having many smart contracts. 
Figure \ref{fig:cdf_sc} shows that 70.57\% of dapps have just one smart contract, and 90.46\% of dapps have less than 3 smart contracts. But there are some dapps having more than 100 smart contracts.

\begin{figure}[!t]
    \centering
    \includegraphics[width=2.5in]{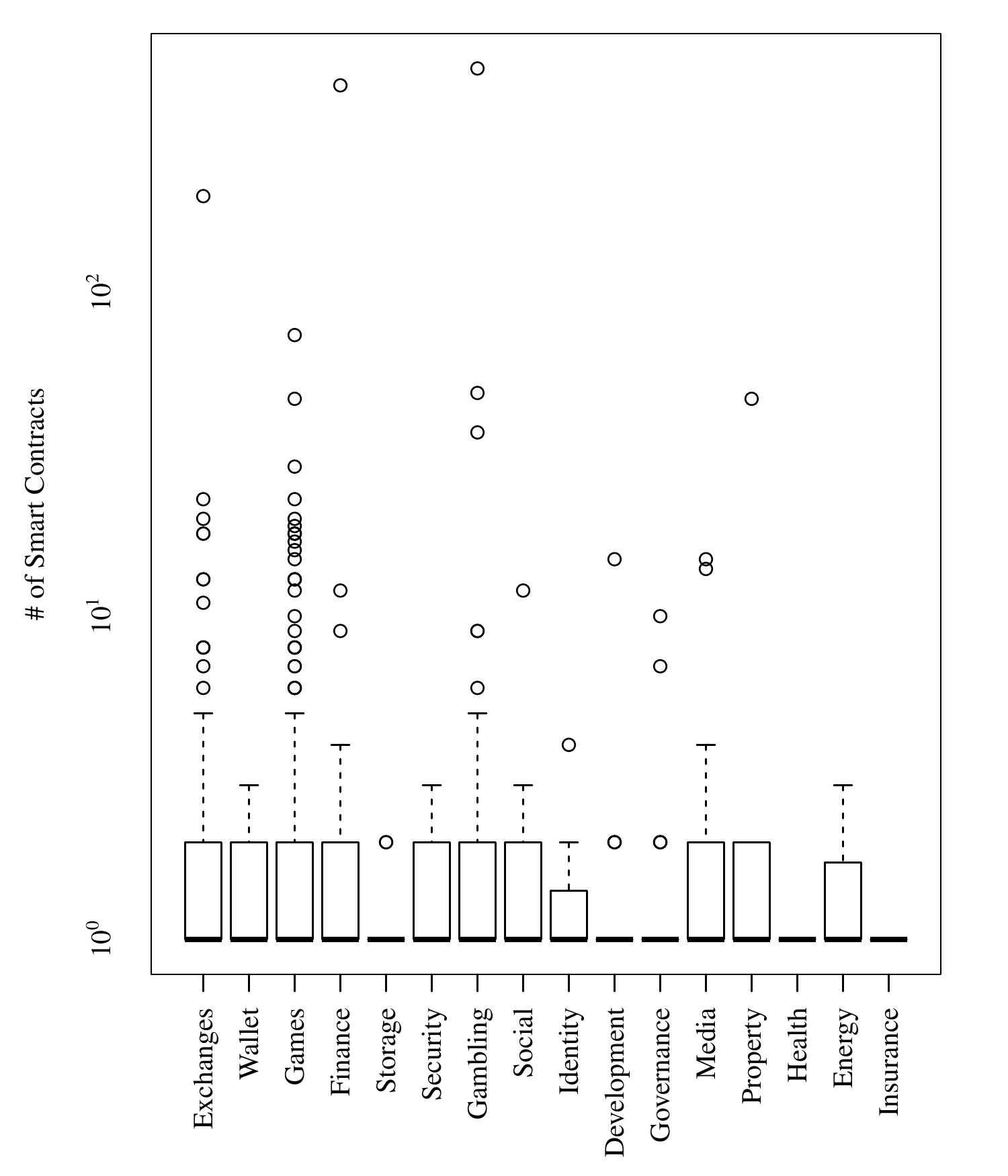}
    \caption{Distribution of the numbers of smart contracts by Categories}
    \label{fig:cat_sc_box}
\end{figure}

Dapps of different categories have different design objectives, so that there is difference among numbers of smart contracts of dapps of different categories. In figure \ref{fig:cat_sc_box}, boxes from left to right represent 16 categories respectively shown in Table \ref{tab:categories}. We can find that category Exchanges, Games, Finance and Gambling have more dapps with a large number of smart contracts.
We will discuss about these dapps and smart contracts in Section \ref{sec:reuse}.

\subsection{Distribution by LoC}

\begin{figure}[!t]
    \centering
    \includegraphics[width=2.5in]{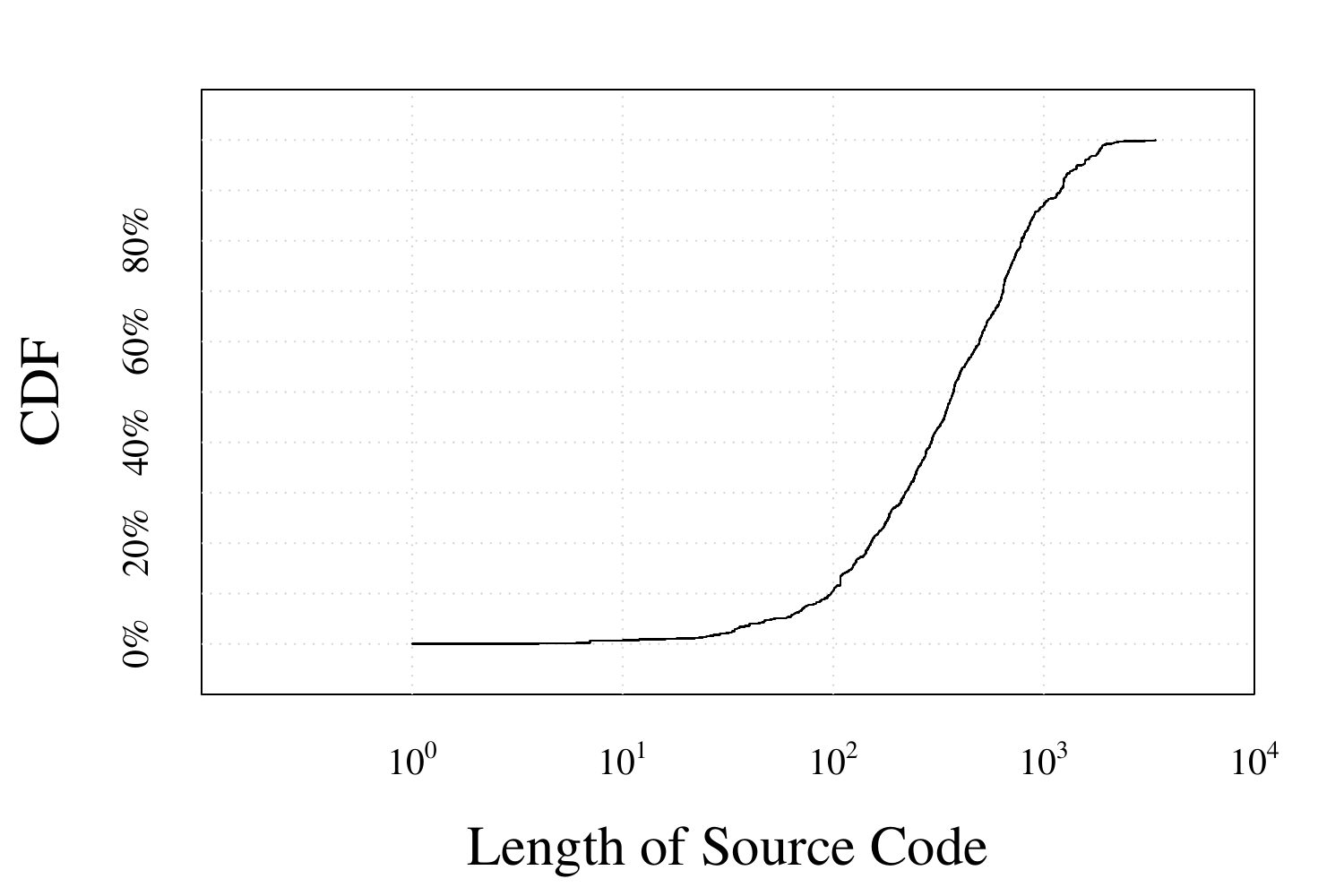}
    \caption{Length of Source Code of a Dapp}
    \label{fig:distribution_loc}
\end{figure}

After removing empty and duplicate source codes, figure \ref{fig:distribution_loc} shows the distribution of dapps by LoC. It shows that length of codes of 87\% of smart contracts is less than 1,000. Source codes of about 70\% dapps' smart contracts have hundreds of lines.

\begin{figure}[!t]
    \centering
    \includegraphics[width=2.5in]{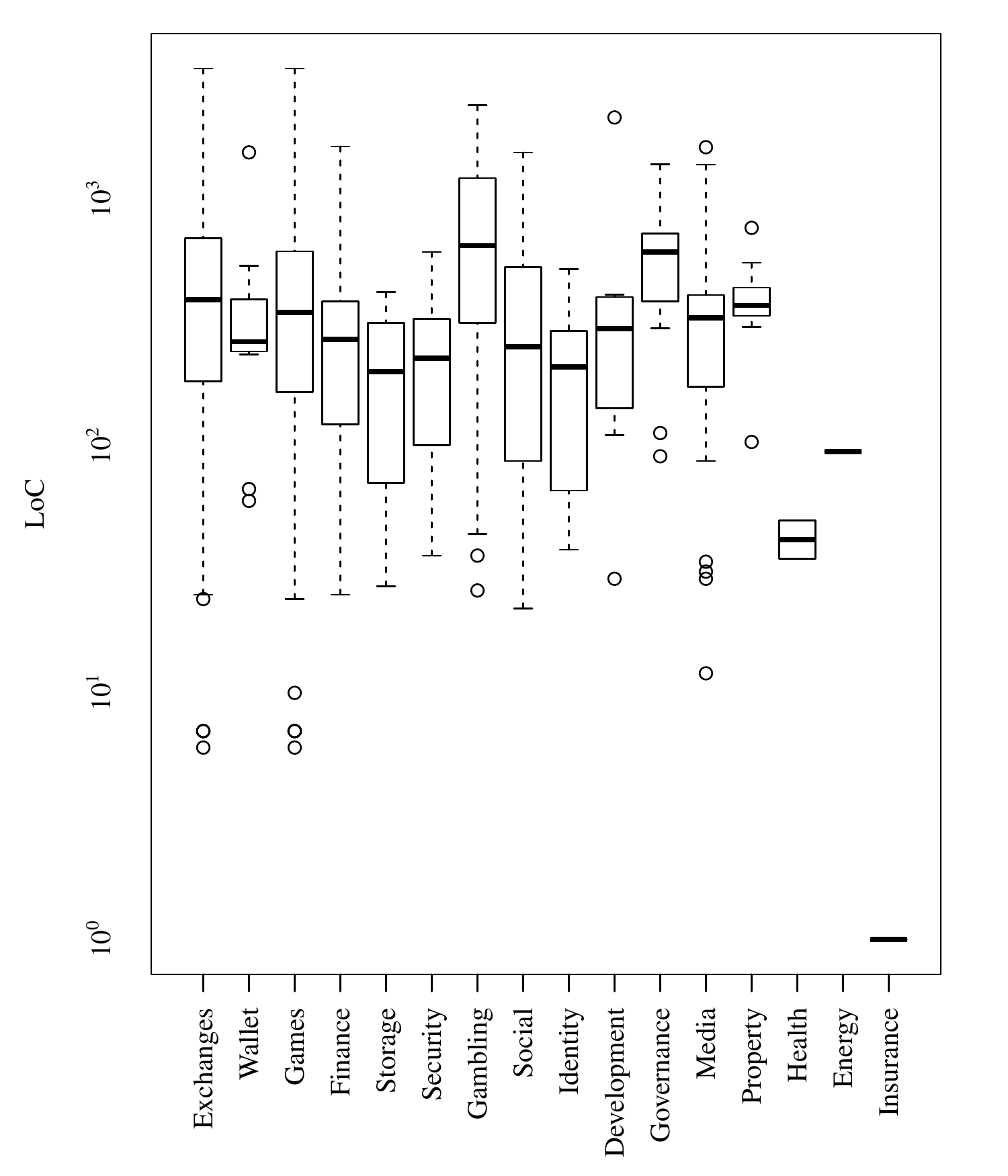}
    \caption{LoC Distribution by Categories}
    \label{fig:cat_loc_box}
\end{figure}

LoC distribution by categories is shown in figure \ref{fig:cat_loc_box}. Generally, LoC of dapps of category Gambling and Governance is higher than the others', because it needs more codes to implement additional mechanisms, such as principles of games, kind matching and secure payment.

\subsection{Reuse of Smart Contracts and Source Codes}
\label{sec:reuse}

Smart Contracts are immutable after deployment. If developers want to modify or remove it, they need publish a new version and make the older one destruct itself. Because of the difficulty of data migration, developers can reuse the old, or abandon data kept by old contracts. So there are two types of reuse of smart contracts: contract-level reuse and code-level reuse.

\subsubsection{Contract-level Reuse}

\begin{table}[!t]
    \caption{Contract-level Reuse in Dapps}
    \centering
    \begin{tabular}{lr}
        \hline
        {\bf \small Dapp Title} & {\bf \small Number of Common Contracts} \\
        \hline
        CryptoLotto & \multirow{2}{*}{9}\\
        CryptoLotto Official & \\
        \hline
        Crypto Sportz & \multirow{2}{*}{1} \\
        SportCrypt & \\
        \hline
    \end{tabular}
    \label{tab:con_level_reuse}
\end{table}

Contract-level reuse, namely different dapps use same deployed smart contracts, can be checked by comparing contract addresses. We find 10 addresses of 2 dapps are used twice, shown in Table \ref{tab:con_level_reuse}. The two dapps all belong to category Gambling, each game is controlled by a smart contract. The contract-level reuse is just re-registering games in new dapp.

\subsubsection{Code-level Reuse}

\begin{table}[!t]
    \caption{Code-level Reuse in Dapps}
    \centering
    \begin{tabular}{lrr}
        \hline
        {\bf \small Category} & {\bf \small Nonempty Source Codes} & {\bf \small Percentage} \\
        \hline
        Exchanges & 420 & 48.10\% \\
        Wallet & 13 & 0\% \\
        Games & 551 & 8.71\% \\
        Finance & 75 & 4\% \\
        Storage & 15 & 0\% \\
        Security & 13 & 0\% \\
        Gambling & 642 & 71.96\% \\
        Social & 62 & 1.61\% \\
        Identity & 9 & 0\% \\
        Development & 10 & 30\% \\
        Governance & 19 & 0\% \\
        Media & 56 & 3.57\% \\
        Property & 15 & 26.67\% \\
        Health & 2 & 0\% \\
        Energy & 1 & 0\% \\
        Insurance & 0 & NaN \\
        All dapps & 1,597 & 55.48\% \\
        \hline
    \end{tabular}
    \label{tab:code_level_reuse}
\end{table}

To find code-level reuse in dapps, we extract md5 from smart contract source codes, and compare them. Just 1,597 of 2,540 smart contracts provide source codes. We extract 886 unique md5 value of these source codes. Code-level reuse in dapps of each category is shown in Table \ref{tab:code_level_reuse}. We can find that category Gambling has the highest percentage of code-level reuse. Because to manage a game in Gambling just a smart contract needs to be deployed, smart contracts with the same source codes are deployed again and again. Category Exchanges has the second highest percentage. There are many active smart contracts with the same source codes in this category, deployed to handle similar requests. It doesn't help to improve throughput of dapps, but is useful for managing data of different kinds. We can find the similar usage in other dapps.

Surprisingly, there are two smart contracts of two dapps have the same source code: dapp Franklin's Library and Integrity Warriors. The former one belongs to category Games with the latter one of category Gambling. They have similar numbers of unique users, transactions and transaction volume, and the latter one is marked as "Ponzi schemes" by \textit{State of the ÐApps}.

\subsection{Case Study}

\begin{table}[!t]
    \caption{Information of Ether Goo}
    \label{tab:dapp_info}
    \centering
    \begin{tabular}{ll}
        \hline
        {\bf \small Website} & https://ethergoo.io/ \\
        {\bf \small Open Source} & Open source smart contracts \\
        {\bf \small Number of Smart Contracts} & 3 \\
         & 1113 \\
        {\bf \small LoC} & 892 \\
         & 142 \\
        {\bf \small Unique Users in the year} & 6,643 \\
        {\bf \small Transactions in the year} & 312,140 \\
        {\bf \small Transaction Volume} & About 1127.21 ETH \\
        \hline
    \end{tabular}
\end{table}

We select a hot dapp of category Games, Ether Goo \cite{sodapp_ether_goo}, to do case study. Ether Goo is an idle game, in which players just need to set some configurations and then wait for characters or organizations to work. But if players pay ethers, they will get ahead of other players, which means they will have some advantages like useful but rare items and higher speed of upgrading. Authors call it the first competitive idle game. This dapp was submitted to \textit{State of the ÐApps} at Mar 28, 2018. Summary of this dapp and its related information is shown in Table \ref{tab:dapp_info}.

\begin{figure}[!t]
    \centering
    \includegraphics[width=2.5in]{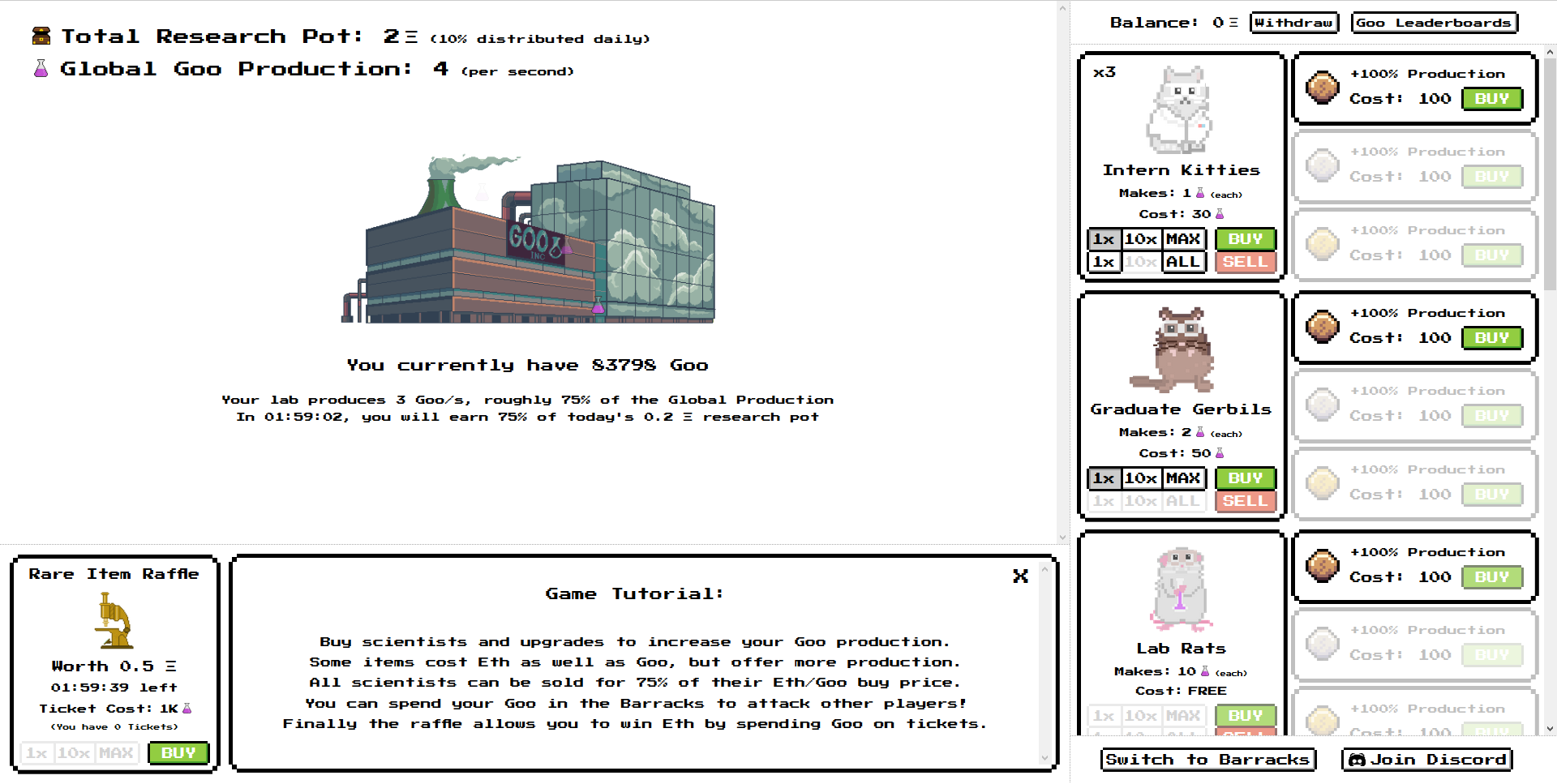}
    \caption{A Screenshot of Ether Goo}
    \label{fig:ether_goo}
\end{figure}

Ether Goo consists of three parts, a web UI, a back-end and three smart contracts.
The web UI \cite{ether_goo} is used to interact with users and show status of their characters and other things, like figure \ref{fig:ether_goo}.
We can infer from source codes of web UI and smart contracts that the back-end receives requests from the web UI and then transform them to transactions on Ethereum to call smart contracts, and saves some data as well. 

In smart contracts, two of them are used to issue tokens following \textit{ERC-20 Standard} (a standard of implementing tokens on Ethereum) \cite{erc20}, and handle key operations of the game. They have similar source codes, but the contract with shorter codes had few transactions four days after the day at which the longer one was submitted. It means that authors used to upgrade key operations. Because of issuing tokens, the older smart contract \cite{goo_older_key_contract} was not been destructed and not used to support the game any more. We call the newer smart contract key smart contract \cite{goo_newer_key_contract}.

The source code of key smart contract consists of three parts: an interface, two contracts and a library. The interface works like interface in Java, named as \textit{ERC20}, in which there are some functions required in \textit{ERC-20 Standard}. Contracts, \textit{Goo} and \textit{GooGameConfig}, works like class in Java, used to keep data and support operations. Besides functions interface \textit{ERC20} requires, \textit{Goo} also defines functions to support the game, such as function \textit{getGameInfo()} and \textit{attackPlayer()}.

Another smart contract \textit{GooLaunchPromotion} \cite{goo_promotion_contract} is used to do a promotion. Users deposited ethers to the smart contract and user with the highest ethers deposited would got some prizes. Ethers deposited were sent to users after awarding prizes. The promotion started at Mar 27, 2018 \cite{goo_promotion_start} and ended at Mar 31, 2018 \cite{goo_promotion_end}.

%% file: secs/findings.tex
\section{Summary of Findings}
\label{sec:findings}


From analysis above, we can get some conclusions.

First, the dapp popularity generally complies with the Pareto principle in terms of unique users, transactions and transaction volumes, and access time. A "long-tail" of dapps are rarely active and have high percentage.

Second, Exchanges, Wallet and Games are the hottest three categories. Category Exchanges and Wallet have almost all of transaction volume. The number of dapps of category Health, Energy and Insurance grows slowly and their LoCs are lower, it means new use of dapps is still in exploration.

Third, open source smart contracts make dapps more popular, but it doesn't have obvious influence to make other codes open source.

Finally, code-level reuse in dapps is common. In Gambling, the reuse is to start a new game. And in other categories, the reuse is usually to handle similar but different transactions, e.g. two kinds of tokens generated by one source code.

%% file: secs/implications.tex
\section{Implications}
\label{sec:implications}

We have investigated popularity and smart contracts of dapps and get some conclusions. In this sections, we suggest some implications.

First, for dapp collecting website owners, a few dapps that have high percentage of unique users, transactions or transaction volumes are main web traffic entrances. Other dapps can be added to related list of these dapps to get more attention. It's an effective way to advertise, and website owners benefit from it.

Second, it is important to decide what category the dapp belongs to before publishment. It doesn't mean that the functions of the dapp change by categories, developers just need to decide which features the dapp shows for users and dapp collecting website owners. The hotter category is, the longer "long-tail" of dapps is. But it is difficult to get users for dapps like those of category Health, Energy and Insurance, unless the dapp have unique characteristics or subversive functions. And for some categories, such as Gambling and Media, smart contracts need to be designed for special requirements, e.g. secure payment.

Third, developers would better submit source codes of smart contracts to a open-source website or Ethereum block explorer. It may help the dapp to get trust of users.

Fourth, if the dapp belongs to category Gambling, it doesn't matter that deployed smart contracts are immutable. Usually same source codes are used again and again, so developers can modify the source code any time and deploy it for next game. Meanwhile, users of this kind of dapps have to check often md5 of the source code to prevent financial loss.
For other developers, it is useful to deploy a same source code times to handle similar but different requests, especially when there are lots of kinds of requests.

Finally, any apps with related payments can be transferred to dapp version. Through smart contracts, it is easy to receive ETH from users. Developers can also issue tokens to get foundations. If developers want smart contracts to handle more core operations, there are some guiding principles:

\begin{itemize}

    \item What data does smart contracts have?

    The data that smart contracts easily keep is the token-related. Making smart contracts hold other data is OK but be care of limitations of smart contracts: smart contracts can't hold much data because of small room on chain.

    \item What functions does the dapp have?

    Developers have to code for function requirements. It costs gas to run program on Ethereum blockchain, so length of function codes must be suitable. If a function is too long there will be too many calling failures because the gas users send isn't usually enough.

    \item How many smart contracts does the dapp have?

    Note that deploying a smart contract also costs gas. Smart contracts with too long source code can't be deployed, for used gas is over the gas limit of a block. Split smart contracts properly to avoid it.

\end{itemize}

Nowadays, the most usual apps transferred to dapps are gambling games, collectible card games and apps with internal payments.

%% file: secs/relatedwork.tex
\section{Related Work}
\label{sec:related_work}

Blockchain has become a hot research field. Researchers mainly focus on three topics: underlying mechanisms, applications and data mining.
Underlying Mechanisms include consensus mechanisms and smart contract mechanisms. Many consensus mechanisms are proposed, such as PoW (Proof of Work) \cite{nakamoto2008bitcoin}, PoS (Proof of Stake) \cite{king2017ppcoin}, DPoS (Delegated Proof of Stake) \cite{larimer2014delegated}, and Algorand \cite{gilad2017algorand} recently presented. Some classical consistency algorithms like PBFT \cite{barger2017scalable} (practical byzantine fault tolerance) are used as well. Furthermore, a few researchers try to measure and improve existing mechanisms \cite{zhang2017rem, zheng2018detailed}. Smart contract mechanisms attract experts of software engineering \cite{liao2017toward, porru2017blockchain} and security \cite{coblenz2017obsidian}.
For its decentralization, persistency, anonymity and auditability, blockchain can be used in anonymous trading, persistence services and cross-organizational transactions. So blockchain technology has been widely used in finance service \cite{klems2017trustless}, IoT (Internet of Things) \cite{shae2017design, ruta2017supply, bahri2018trust}, information security \cite{liu2017blockchain, Nguyen:2018:TAT:3184558.3186938} and software engineering \cite{liao2017toward}.
Thanks to the public accessibility and auditability of blockchain, researchers can do analysis based on transaction data for usage characteristics \cite{ron2013quantitative, meiklejohn2013fistful} and so on. The other accessible data is smart contract code, for example, bytecode written in the block. Because of the wave of open source, source codes of most of smart contracts can be found in Github or other open website like Etherscan. There are some work analyzing these codes to give advises for smart contract developers and blockchain users \cite{chang2018scompile, chen2018detecting}.

In recent years, mobile web becomes large and accumulate lots of data. There are some several analyzing work based on the data, including usage data \cite{li2015characterizing, lu2016prada, lu2017prado, liu2018understanding} and management data \cite{li2015descriptive, li2016voting, liu2017deriving}. Compared to mobile web, blockchain has similar and more real usage data. Knowledge learned from these data in the same way is useful for understanding blockchain users and dapp market as well.

%% file: secs/conclusion.tex
\section{Conclusion and Future Work}
\label{sec:conclusion}

The scale of dapp market is gradually growing and has reached billions of dollars. In this study, we conducted a systematic descriptive analysis of 734 dapps over Ethereum. We find relationships among dapp popularity, categories, open source and LoC, and explore reuse of smart contracts and source codes. Our findings provide valuable implications for different stakeholders in the decentralized application industry and in the research community of blockchain and decentralized applications.

This paper mainly focuses on the descriptive analysis of the dapps. In future, we will focus questions about dapp projects and source codes, such as how to build a dapp project, how to keep synchronization on and off the chain, and how to improve throughput of dapps. These researches can directly improve the development of dapps and benefit millions of users.